\documentclass[]{article}
\usepackage{amsmath,amssymb}
\usepackage{graphicx}
\usepackage{dcolumn,natbib}
\newcommand{\mbf}[1]{\mathbf{#1}}
\newcommand{\llangle}[0]{\langle\!\langle}
\newcommand{\rrangle}[0]{\rangle\!\rangle}

\title{Generalized optical theorems for the  reconstruction of  Green's function of an inhomogeneous elastic medium}
\author{Ludovic Margerin, Dynamique Terrestre et Plan\'etaire,  \\ Observatoire Midi-Pyr\'en\'ees, C.N.R.S.,
  \\ 14 Avenue Edouard Belin, Toulouse,  France \and
Haruo Sato, Department of Geophysics, \\ Graduate School of Science, Tohoku University, \\ Aramaki-Aza Aoba 6-3, Aoba-ku, Sendai, Miyagi, 980-8578 JAPAN}
\begin{document}
\maketitle
 \begin{abstract}
  This paper investigates the reconstruction of elastic Green's function from the cross-correlation  of waves excited by  random noise  in the context of scattering theory. Using a general operator equation, -the resolvent formula-,  Green's function reconstruction is established when the noise sources satisfy an equipartition condition. In an inhomogeneous medium, the operator formalism leads to generalized forms of  optical theorem involving the off-shell $T$-matrix of  elastic waves,  which  describes scattering in the near-field. The role of  temporal absorption in the formulation of the theorem is discussed. Previously established symmetry and reciprocity relations involving the on-shell $T$-matrix are recovered in the usual far-field and infinitesimal absorption limits. The theory is applied to a point scattering model for elastic waves. The $T$-matrix of the point scatterer incorporating all recurrent scattering loops is obtained by a regularization procedure. The physical significance of the point scatterer  is discussed. In particular this model satisfies the off-shell version of the generalized optical theorem. The link between equipartition  and Green's function reconstruction in a scattering medium is discussed.
 \end{abstract}
\maketitle
\section{Introduction}
Recently, there has been increasing interest in the use of coda and scattered waves to monitor temporal
variations in dynamic structures such as volcanoes and faults. Such an approach is known as coda-wave interferometry
in seismology and diffusive wave spectroscopy in acoustics \citep{snieder2002,sniederpage2007}.  While the early developments of the
technique were based on the use of repeating small earthquakes termed ``doublets'' \citep{poupinet1984,poupinet2008}, the possibility
to reconstruct Green's function from ambient seismic noise  has given rise to a new technique termed passive image interferometry
 \citep{SensSchoenfelder.Wegler.2006,wegler2007,Brenguieretal.2008b,Brenguieretal.2008a}.
 In  coda-wave interferometry, temporal changes in the medium are detected by comparing the coda portions 
of the cross-correlation function of ambient noise, that were computed at different times. 
Therefore, understanding the basic physical processes that allow the reconstruction
of scattered waves from the cross-correlation of random noise sources is of interest in seismology
and acoustics.

 The reconstruction of Green's function from field cross-correlations  has a rich and vast  history as reviewed by \cite{shapiro2011}.
 Green's function retrieval in seismology and elastodynamics finds its roots in the fluctuation-dissipation theorem and in the theory of speckle correlations in optics.
 In recent years, the pioneering results described by  \cite{shapiro2011}  have been extended to different system configurations:  
open or closed, and to  different types of excitations: deterministic or
 random sources located on a surface or distributed in a volume. For instance, general theorems on the 
reconstruction of Green's function  have  been established for a variety of inhomogeneous, dissipative  \citep{Wapenaar2006,sniederwapenaar2007},
 and even non-reciprocal media  \citep{Wapenaar.2006}.
 An important ingredient of seismic interferometry is Green's theorem or its
elastodynamic generalization, i.e. the Rayleigh-Betti reciprocity theorem \citep{ramirez2009}. The importance of Green's theorem and its generalizations
for the reconstruction of Green's function is  well illustrated by \cite{Wapenaar.Fokkema.2006}. Depending on the excitation mechanism -active or passive-,
the acquisition geometry -surface or borehole measurements-, the nature of boundary conditions -radiation or reflection-, the type of Green's function employed
-analytical, numerical, empirical-, a great variety of interferometric methods can be derived from the application of Green's theorem. A comprehensive
review of applications to exploration seismology, emphasizing the strengths and limitations of each method is provided by \cite{ramirez2009}. 

 More specifically,
 the connection between the cross-correlation of  wavefields  and Green's function
  between two points in an inhomogeneous elastic half-space has been established by \cite{wapenaar2004},  based on Betti's reciprocity theorem.
 The reconstruction of Green's function of a homogeneous
elastic medium has been discussed by \cite{SanchezSesma.Campillo.2006} from a different perspective. 
These authors find that a set of uncorrelated plane $P$ and $S$ waves
verifying the equipartition relation, i.e. carrying the same amount of energy, allow the elastic Green's function to be reconstructed
from the cross-correlation of noise  wavefields.
This result was extended to the  case of one cylindrical inclusion in a 2-D elastic medium  by \cite{sanchezsesma2006b}. These authors
demonstrated the complete reconstruction of Green's function including scattered waves in the case of an illumination of the heterogeneous medium
by an equipartition mixture of plane $P$ and $S$ waves.  

The case of scattering media has been studied by \cite{Sato.2009a,Sato2010}. He demonstrated the reconstruction of singly-scattered coda waves 
from random noise sources in an acoustic medium composed of point-like velocity perturbations. He considered both volumetric and surface distribution
of noise sources.
An important connection between energy conservation in scattering and the reconstruction of scattered waves in
 Green's function obtained from cross-correlations of random noise sources has recently been  put forward by \cite{Snieder.2008,Sniederetal.2009}.
Considering a scattering medium illuminated by  noise sources distributed on an enclosing surface,  \cite{Snieder.2008} showed that the 
reconstruction of  Green's function of scalar waves is equivalent to a generalized optical theorem. Such theorems provide general
symmetry relations that must be obeyed by the scattering amplitude, which describes the amplitude and phase relations between
incident and scattered waves.   \cite{lu2011}  extended the method of \cite{Snieder.2008} to the case  
of  vector elastic waves,  for a scattering medium composed of a homogeneous and isotropic heterogeneity.
  \cite{Snieder.Fleury.2010} and  \cite{margerin2010}  developed a theory for the reconstruction of scattered arrivals
in  Green's function obtained from the cross-correlation of noise signals in an acoustic medium, for surface and volumetric
distributions of noise sources, respectively. These authors  demonstrate the cancellation of spurious terms in the multiple scattering series expansion of 
Green's function cross-correlations by repeated application of the optical theorem.

In this work, we analyze  the case of a general
-possibly anisotropic- heterogeneity embedded in an isotropic medium, illuminated by a volumetric distribution of random, uncorrelated forces.
 In the same spirit as \cite{Snieder.2008}, \cite{lu2011}, and \cite{margerin2010}, we establish
the equivalence between Green's function reconstruction and some symmetry relations pertaining to the scattering of elastic waves
which we refer to as ``generalized optical theorems''. Compared to our previous work on scalar waves, we first
use a general operator formalism to establish in a concise way  Green's function reconstruction theorem for a general -homogeneous or
inhomogeneous- elastic medium. Next, we apply the theorem to Green's function of a scattering medium to establish general symmetry
relations in the scattering process. An analysis based on scattering theory finally shows that the symmetry relations are extensions
of generalized forms of the optical theorem for elastic waves. The central result of this paper -Equation (61)- contains as special cases 
all forms of optical theorems found in the literature.
  In view of future applications to a collection of scatterers we develop
a simple point-like scattering model for elastic waves.  The formulas (\ref{top})-(\ref{tmat}) present an extension of the
 point-scattering model originally developed for scalar waves to the case of vector elastic waves.
 The  physical interpretation of the model is discussed in details. 
\begin{figure}
\centering
\includegraphics[width=12.5cm]{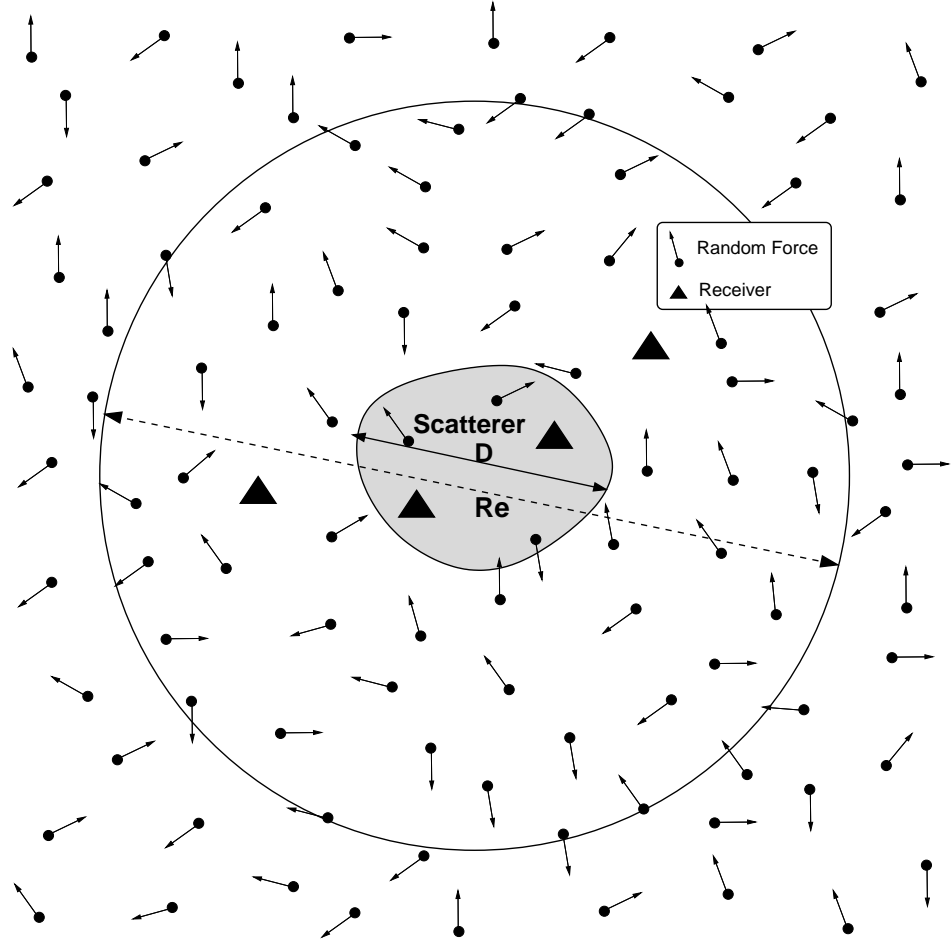}
\caption{Problem setting. We consider the reconstruction of  Green's function including the scattered waves
 from the cross-correlation of random wavefields recorded  in a homogeneous isotropic elastic medium containing 
an arbitrary  heterogeneity of typical dimension $D$ (grey region). Receivers may be placed
at arbitrary positions in the medium, including the near-field or the interior of the scatterer.
The  applied forces   are stationary in time, 
randomly and uniformly oriented, and randomly and homogeneously distributed.
The effective radius of the sources is  $R_e = c_s\tau $ with
$\tau$ the absorption time and $c_s$ the shear wavespeed. The effective radius is (much) larger than the size
of the heterogeneity $D$ and the typical distance between the receivers.}
\label{pbsetting}
\end{figure}

The problem setting is depicted in Figure \ref{pbsetting}. We consider an arbitrary  heterogeneity with bounded support
embedded in an elastic medium illuminated by a randomly homogeneous distribution of  forces oriented in all possible
space directions. By considering the correlation of the  displacements recorded at two points $A$
and $B$, we will show the possibility to reconstruct the elastic Green's tensor including the waves
scattered by the heterogeneity.
In a heterogeneous, slightly dissipative elastic medium, the wavefield generated by random forces $\mathbf{f}(\mathbf{x},t)$ 
satisfies the following equation:
\begin{multline}
 \rho_0 \left( \dfrac{\partial^2 u_i(\mathbf{x},t)}{\partial t^2} + \dfrac{1}{\tau} \dfrac{\partial u_i(\mathbf{x},t)}{\partial t} \right) 
   + \delta \rho(\mathbf{x}) \dfrac{\partial^2 u_i(\mathbf{x},t)}{\partial t^2}   
-\partial_{x_j} C^0_{ijkl} \partial_{x_k} u_l(\mathbf{x},t)  \\  -\partial_{x_j} \delta C_{ijkl} \partial_{x_k} u_l(\mathbf{x},t)    = f_i(\mathbf{x},t),
\label{elasto1}
\end{multline}
where $\rho_0$ and $C^0_{ijkl} = \lambda_0 \delta_{ij}\delta_{kl} + \mu_0(\delta_{ik} \delta_{jl} + \delta_{il}\delta_{jk} ) $ denote the
density and elasticity tensor of the isotropic matrix. The summation over repeated indices is assumed.
The second term on the left-hand side introduces a characteristic absorption time
 $\tau$. The deviations of the density and  elastic properties from homogeneity are encapsulated in the terms
 $\delta \rho = \rho - \rho_0$ and $\delta C_{ijkl} = C_{ijkl} - C^0_{ijkl}$. No particular symmetry properties, except the usual ones, are imposed 
on $C_{ijkl}$. On the right-hand side of equation (\ref{elasto1}), we consider a  randomly homogeneous and 
stationary distribution of forces with uniform random orientations. In addition, we assume that the spatial correlation
of these forces is much shorter than any other scale length of the problem, which implies that they are well described
by a white noise process in space.  The basic observable in seismology is the cross-correlation of signals averaged over an ensemble of noise sources:
\begin{equation}
 C_{ij}(\mathbf{x}_B,\mathbf{x}_A;t) = \lim_{T \to +\infty}  \dfrac{1}{T} \int^{T/2}_{-T/2} \llangle  u_i(\mathbf{x}_B;t') u_j(\mathbf{x}_A;t'-t)^* \rrangle dt' ,
\label{crosscorrel}
\end{equation}
where the double brackets $\llangle \cdot \rrangle$ denote the ensemble average. 
If the noise sources are statistically stationary, the basic quantity to be evaluated is the integrand 
on the right-hand side of  equation (\ref{crosscorrel}).
 Note that the definition of the cross-correlation adopted in Eq. (\ref{crosscorrel}) differs from that found in 
other references \citep[see e.g.][]{wapenaar2004}. For deterministic or stationary random signals, the change of variable
 $t'\rightarrow t'+t$ allows us to recover the form of cross-correlation used in other references. 
The difference is therefore purely notational.
 Since calculations are much easier to perform in the frequency domain, we will develop a theory for the cross-spectral density:
 \begin{equation}
   C_{ij}(\mathbf{x}_B;\mathbf{x}_A;\omega)  =  \int^{\infty}_{-\infty}    \llangle  u_i(\mathbf{x}_B;t') u_j(\mathbf{x}_A;t'-t)^* \rrangle e^{i\omega t} dt
\label{crosscorrelfreq}  
\end{equation}
In equation (\ref{crosscorrelfreq}), the left-hand side does not depend on the variable $t'$ for a stationary signal. 
Upon introducing the spectral representation of the random process $u_i(\mathbf{x},t)$:
\begin{equation}
 u_i(\mathbf{x};t) = \dfrac{1}{2\pi} \int_{-\infty}^{+\infty} u_{i}(\mathbf{x};\omega) e^{-i \omega t} d\omega,
\label{uix}
\end{equation} 
we obtain the following basic relation:
\begin{equation}
  \llangle u_i(\mathbf{x}_B;\omega') u_j(\mathbf{x}_A;\omega)^* \rrangle = 2\pi C_{ij}(\mathbf{x}_B,\mathbf{x}_A;\omega) 
   \delta(\omega-\omega').
 \label{fieldcorstat}
\end{equation}
The delta function condition which appears in equation (\ref{uix}) is characteristic of  random processes that are stationary in time. 
We note that  the usual Fourier transforms $u_i(\mathbf{x};\omega)$ may not be well defined since we are interested in 
stationary random fields.  A physically satisfying solution is to define the Fourier transforms for a finite observation time 
$T$, and to take the limit $T \to \infty$  after averaging over noise sources. 
Instead, we make use of the  probabilistic interpretation of the spectral representation  (\ref{uix}),
where the coefficients of the exponential  is itself a random process, and
 the equality  is to be understood in the probabilistic (mean-squared) sense \citep{rytov1989t3, yaglom2004}. 
Since only the cross-spectral density of the field is needed in this work, and this quantity can be defined by traditional Fourier
 analysis,  the formal representation (\ref{uix}) suffices. 
A link between the physical and mathematical approaches
is provided by the periodogram:
\begin{equation}
  C_{ij}(\mathbf{x}_B,\mathbf{x}_A;\omega) = 
        \lim_{T \to \infty}     \dfrac{1}{T} \llangle  u^T_i(\mathbf{x}_B;\omega)  u^T_j(\mathbf{x}_A;\omega)^*   \rrangle,
\label{periodo}
\end{equation}
where  $u_i^T(\mathbf{x};\omega)$ represents the Fourier transform of the original signal observed over a finite time window $T$. 
As is well-known, Equation (\ref{periodo}) expresses the fact that the periodogram is an asymptotically unbiased estimator
of the cross-spectral density.

The paper is organized as follows. In section \ref{formalism}, we introduce some important notations and 
 present a simple derivation of  Green's function reconstruction
for a homogeneous medium illuminated by a set of randomly oriented and spatially uncorrelated forces. In section \ref{formalism2}
we introduce the Dirac calculus which will be used to generalize our results to an inhomogeneous medium. The formalism is applied to demonstrate
the reconstruction of  Green's function of a general -homogeneous or inhomogeneous- elastic medium illuminated by 
 a homogeneous distribution of randomly oriented forces,  or by a wavefield at equipartition. 
In section \ref{scattering}, we introduce
the $T$-operator for elastic waves and discuss its symmetry properties.
 In section \ref{general}, we explore  the reconstruction of  Green's function
for an arbitrary  inhomogeneity embedded in a homogeneous medium and establish several forms of the generalized
optical theorem for elastic waves. 
 In section \ref{formalproof}, we establish the generalized optical theorem from the governing equations of scattering
 in the limit of infinitesimal absorption. In section \ref{point}, we consider
a simple configuration composed of a single point scatterer and demonstrate the Green's function reconstruction including
all the recurrent scattering loops. In section \ref{conclusion}, our results are  discussed in the framework of
 equipartition theory and compared to other works.

\section{Green's function reconstruction in a  homogeneous medium}
\label{formalism}
In this section, we introduce important notations and present in a simplified context our approach to Green's function reconstruction.
Although it is not the main theme of the paper, we give a  short  derivation of  Green's function reconstruction for elastic waves excited 
by random uncorrelated noise sources in a homogeneous medium. Our assumptions differ from the one adopted by \cite{SanchezSesma.Campillo.2006} who considered
 a set of uncorrelated plane $P$ and $S$ waves
at equipartition. The complete equivalence between the two models of noise source  will be demonstrated later in the paper.
Let us first rewrite the correlation function (\ref{crosscorrelfreq}) in terms of  Green's function.
The  elastic wavefields recorded at two points  $ \mathbf{x}_A$ and $ \mathbf{x}_B$ due to a distribution of forces can be expressed as:
\begin{subequations}
\begin{align}
u_i(\mathbf{x}_B;\omega') = & -\iiint_{-\infty}^{\infty} G^0_{ik}( \mathbf{x}_B,\mathbf{x}_0;\omega') f_k(\mathbf{x}_0;\omega') d^3x_0, \\
u_j(\mathbf{x}_A;\omega)^* = & - \iiint^{\infty}_{-\infty} G^0_{jl}(\mathbf{x}_A,\mathbf{x}_1;\omega)^* f_l(\mathbf{x}_1;\omega)^* d^3x_1,
\end{align} \label{uia}
\end{subequations}
where $G^0(\omega)$ is  Green's function of a homogeneous, slightly dissipative
elastic medium  at angular frequency $\omega$. 
Using the summation convention over repeated indices, $G^0(\omega)$ is recognized as the fundamental solution  to the equation of elastodynamics:
\begin{equation}
   \left( \mu_0 \partial{x_l}\partial{x_l} \delta_{ij}  + (\lambda_0 + \mu_0)\partial_{x_i} \partial_{x_j} \right) G^0_{jk}(\boldsymbol{x},\boldsymbol{x}';\omega)   
 + \rho_0 (\omega + i/2\tau)^2 G^0_{ik}(\boldsymbol{x},\boldsymbol{x}';\omega) = 
 \delta_{ik} \delta(\boldsymbol{x},\boldsymbol{x}') 
\label{green1}
\end{equation}
and  has the following analytical form:
\begin{equation}
  G^0_{ik}(\mathbf{x},\mathbf{x}';\omega)  =  \left( g^{s}(\mathbf{r};\omega) +  g^n(\mathbf{r};\omega) \right)   (\delta_{ik} - \hat{r}_i \hat{r}_k)  +
    \left(  g^{p}(\mathbf{r};\omega) - 2 g^n(\mathbf{r};\omega) \right) \hat{r}_i \hat{r}_k , 
\label{g0}
\end{equation}
where  Green's function has been split into transverse and longitudinal parts.
 In equation (\ref{g0}) we have introduced the notation $\mathbf{r} = \mathbf{x} - \mathbf{x}'$,
 the unit vector $\mathbf{\hat{r}}$ in the direction of $\mathbf{r}$,  
 and the functions $g^p$, $g^s$ and $g^n$:
\begin{subequations}
\begin{align}
 g^{s}(\mathbf{r};\omega) = & -\frac{e^{i k_s r}}{4 \pi r \rho_0 c_s^2}  \label{gsrom} \\
   g^{p}(\mathbf{r};\omega) =  &  - \frac{e^{i k_p r}}{4 \pi r \rho_0 c_p^2}  \label{gprom} \\
   g^n(\mathbf{r};\omega)   = &  \frac{1}{4 \pi r^2 \rho_0 }        \left( \frac{i k_p e^{i k_p r}- i k_s e^{i k_s r}  }{\omega_+^2}    
             + \frac{1}{r \omega_+^2} \left( e^{i k_s r} - e^{i k_p r}   \right) \right).
\end{align}
\end{subequations}
 The following short-hand notations have been used: $\omega^+ = \omega + i/2\tau$, $k_p = \omega_+/c_p$ and $k_s=\omega_+/c_s$, where 
$c_p$ and $c_s$ are the $P$ and $S$ wave velocities in the background medium with Lam\'e parameters $\lambda_0$ and $\mu_0$.
The definition of Green's function adopted in equation (\ref{g0}) obeys an opposite
sign convention to that usually found in the seismological literature. This choice will be clarified
in section \ref{formalism2}.
In equation (\ref{g0}), we have explicitly separated the far-field terms $g^s$ and $g^p$ from the near field term $g^n$.  

 Equation (\ref{uia}) allows us to express  the cross-correlation of   wavefields measured
 at $\mbf{x}_A$ and $\mbf{x}_B$ in the frequency domain as follows:
\begin{multline}
\llangle u_i(\mathbf{x}_B;\omega')  u_j(\mathbf{x}_A;\omega)^* \rrangle = \\   \iiint_{\mathbb{R}^6} G^0_{ik}(\mathbf{x}_B,\mathbf{x}_0;\omega') 
G^0_{jk}(\mathbf{x}_A,\mathbf{x}_1;\omega)^*  \llangle f_k(\mathbf{x}_0;\omega') f_l(\mathbf{x}_1;\omega)^*  \rrangle  d^3x_0 d^3x_1
\label{crosscorrelfreq2}
\end{multline}
For spatially uncorrelated and randomly  oriented forces,  one further assumes that
\begin{equation}
\llangle f_k(\mathbf{x}_0;\omega') f_l(\mathbf{x}_1;\omega)^*  \rrangle = 2 \pi \delta_{kl} \delta(\mathbf{x}_0-\mathbf{x}_1) S(\omega) 
  \delta(\omega-\omega') ,
\label{randomforce1}
\end{equation}
where $S(\omega)$ is the power spectral density of the forces.
Therefore the key integral to be computed writes
\begin{equation}
C_{ij}(\mathbf{x}_B,\mathbf{x}_A;\omega) = S(\omega)  \iiint^{\infty}_{-\infty} G^0_{ik}(\mathbf{x}_B,\mathbf{x};\omega) G^0_{jk}(\mathbf{x}_A,\mathbf{x};\omega)^*  d^3x.
\label{keyint}
\end{equation}
 In coordinate space,  Green's function has the complicated form (\ref{g0}). However, using  Fourier transforms,
 it is found that  Green's function has  the following exact expression in wavenumber space:
\begin{equation}
 G^0_{ij}(\mathbf{k}',\mathbf{k};\omega) = 
  g^l(\mathbf{k};\omega) \delta(\mathbf{k}' - \mathbf{k}) \hat{k}_i\hat{k}_j 
   + g^t(\mathbf{k};\omega) \delta(\mathbf{k}' - \mathbf{k}) \left( \delta_{ij} - \hat{k}_i\hat{k}_j \right) 
\label{g0k},
\end{equation}
where we have introduced the following notations:
\begin{align}
  g^l(\mathbf{k};\omega) = & \dfrac{1}{\rho_0(\omega^2 - c^2_pk^2 + i\omega/\tau)},  &
   g^t(\mathbf{k};\omega) = & \dfrac{1}{\rho_0(\omega^2 - c^2_sk^2 + i\omega/\tau)}  ,
\end{align}
and the unit vector $\mathbf{\hat{k}}$ is in the direction of $\mathbf{k}$.
In equation (\ref{g0k}), Green's function has been separated into longitudinal and transverse parts.
Upon inverse Fourier transformation, both contribute to the near-field term in the spatial domain.
In  wavenumber space,  the key integral is defined as
\begin{equation}
\tilde{C}_{ij}(\mathbf{k}',\mathbf{k};\omega)  =  \dfrac{S(\omega)}{(2 \pi)^3} \iiint_{\mathbb{R}^6} C_{ij}(\mathbf{x}_B,\mathbf{x}_A;\omega) 
    e^{-i\mathbf{k}' \cdot \mathbf{x}_B + i\mathbf{k} \cdot \mathbf{x}_A } d^3x_A d^3x_B .
\end{equation}
Inserting the spectral representation of  Green's function in equation (\ref{keyint}), one obtains the following key integral in the
wavenumber  domain:
\begin{equation}
\tilde{C}_{ij}(\mathbf{k}',\mathbf{k};\omega)  = S(\omega)  \iiint^{+\infty}_{-\infty} G^0_{ik}(\mathbf{k}',\mathbf{k}_0;\omega) G^0_{jk}(\mathbf{k},\mathbf{k}_0;\omega)^*  d^3k_0.  \label{C'}
\end{equation}
Reporting expression (\ref{g0k}) into equation (\ref{C'}),
 one obtains:
\begin{equation}
\tilde{C}_{ij}(\mathbf{k}',\mathbf{k};\omega)  =  -\dfrac{S(\omega) \tau}{\rho_0 \omega} \operatorname{Im} g^l(\mathbf{k}) \delta(\mathbf{k}' - \mathbf{k})  \hat{k}_i\hat{k}_j 
      -\dfrac{S(\omega)\tau}{\rho_0 \omega} \operatorname{Im} g^t(\mathbf{k}) \delta(\mathbf{k}' - \mathbf{k}) (\delta_{ij} - \hat{k}_i\hat{k}_j )
\label{g0reconst}
\end{equation}
which establishes the reconstruction of Green's function of a homogeneous, slightly dissipative elastic medium. 
More precisely, equations (\ref{crosscorrelfreq}), (\ref{crosscorrelfreq2}) and  (\ref{g0reconst})  allow us to obtain the following closed form expression
of the cross-correlation function of two wavefields in the space-time domain:
\begin{multline}
\frac{d}{dt} \lim_{T \to \infty } \frac{1}
 {T}\int_{ - T/2}^{T/2}  \llangle u_i\left( {{\boldsymbol{x}}_B ,t'} \right) {u_j\left( {{\boldsymbol{x}}_A ,t' - t } \right)} \rrangle dt' 
  = \\
  \frac{\tau }{2 \rho_0  (2 \pi) }  \int_{ - \infty }^\infty  {d\omega } \left[ {e^{ - i\omega t }
  G^0_{ij}\left( {{\boldsymbol{x}}_B ,{\boldsymbol{x}}_A,\omega  } \right) - e^{ - i\omega \left( { - t } \right)}  
 G^0_{ij}\left( {\boldsymbol{x}}_B ,{\boldsymbol{x}}_A ,\omega   \right)} \right] S\left( \omega  \right).
 \label{gfra}
 \end{multline}
Hence, we retrieve the classical result that the  temporal derivative of the cross-correlation of two wavefields is proportional to
 the difference of the retarded and  advanced Green's functions filtered by the source power spectral density. Strictly speaking, the limit and
 integral signs on the left-hand side of equation (\ref{gfra}) are superfluous. However, introducing the temporal averaging process 
 makes a closer link with the measurement procedure.
It is worth noting that
the result (\ref{gfra}) is valid under the assumption of equal absorption times for $P$ and $S$ waves. We will come back to this point in section \ref{conclusion}.
\section{Green's function reconstruction in an inhomogeneous medium}
\label{formalism2}
\subsection{Formalism and notations}
\begin{figure}
\centering
\includegraphics[width=10cm]{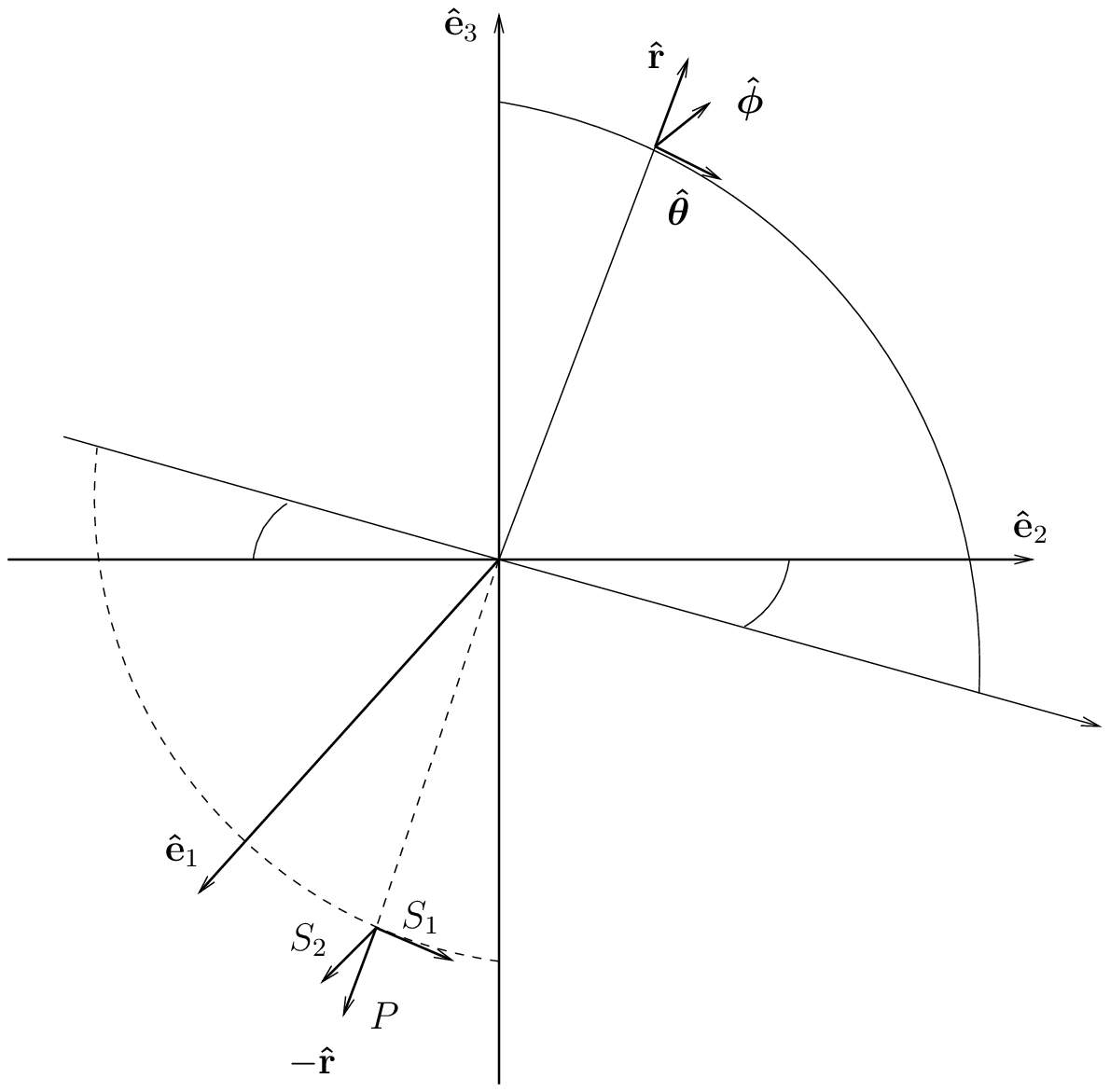}
\caption{Conventions adopted for the definition of the elastic wave polarizations $\left\{S_1,S_2,P\right\}$. $(\mathbf{\hat{e}}_1,\mathbf{\hat{e}}_2,\mathbf{\hat{e}}_3)$ 
denotes a right-handed cartesian system, to which the local right-handed spherical system $(\boldsymbol{\hat{\theta}},\boldsymbol{\hat{\phi}},\mathbf{\hat{r}})$ is referred to. An inversion of the propagation direction $\mathbf{\hat{r}} \rightarrow -\mathbf{\hat{r}}$, implies 
$(\boldsymbol{\hat{\theta}},\boldsymbol{\hat{\phi}}) \rightarrow ( \boldsymbol{\hat{\theta}}, -\boldsymbol{\hat{\phi}}) $.
 The three polarizations $S_1$, $S_2$ and $P$ are shown for the propagation direction $-\mathbf{\hat{r}}$.}
\label{notation}
\end{figure}
To perform the calculations required for the proof of the elastic Green's function reconstruction in an inhomogeneous medium, it is
convenient to introduce the Dirac formalism. This approach allows us to switch easily from
coordinate to wavenumber space and is extremely efficient to perform compact formal derivations.  
In this paper, we work in a  linear space of  vector wavefunctions
in which various representations can be introduced. It is convenient to think
of this vector space as the tensor product of a 3-dimensional polarization space,  and the usual space of scalar wavefunctions. 
 In what follows, the symbol $\langle \mathbf{u} | \mathbf{v} \rangle $ will be employed to denote the scalar
product between two vectors. 
For instance, the space of vector wavefunctions is equipped with the following scalar product:
 \begin{equation}
 \langle \mathbf{u} | \mathbf{v} \rangle  =  \iiint^{+\infty}_{-\infty} u_i(\mathbf{x})^* v_i(\mathbf{x}) d^3x ,
 \end{equation}
which is inherited from the tensor product structure.
 In what follows, we  will switch
among various representations by introducing  generalized orthonormal bases which verify  the continuum normalization.
The three most important bases for this work are listed below.

(1) First, we define:  $| i , \mathbf{x} \rangle = | \mathbf{\hat{e}}_i \rangle \otimes  | \mathbf{x} \rangle $, where the 
symbol $\otimes$ denotes a tensor or  direct product between vectors of the polarization and wavefunction space, respectively.
The bra-ket notations and definitions adopted here follow  the treatment of vector fields in quantum mechanics \citep[see][p.549-551]{messiah1999}.
 The vectors  $ | \mathbf{\hat{e}}_i \rangle$ form an orthonormal  basis of the 3-D polarization space.  The simplest choice is to take 
$|\mathbf{\hat{e}}_i \rangle $ as unit  vectors along the axes of a right-handed orthogonal reference frame. 
We adopt the convention that  unit vectors are  denoted by a hat. 
 The symbol $| \mathbf{x} \rangle$ denotes generalized eigenvectors of  the position operator. By forming the tensor product
 $ | \mathbf{\hat{e}}_i \rangle \otimes  | \mathbf{x} \rangle$, one obtains a wavefunction which is perfectly localized in space at position $\mathbf{x}$
and perfectly polarized along direction $\mathbf{\hat{e}}_i$.
 The completeness of this set of orthonormal vectors  is intuitively clear 
 and the following normalization of the basis vectors is easily verified: 
$ \langle  \mathbf{x} ,i  | j , \mathbf{x}'  \rangle = \langle \mathbf{\hat{e}}_i|  \mathbf{\hat{e}}_j \rangle 
\langle \mathbf{x} | \mathbf{x}' \rangle =  \delta_{ij} \delta(\mathbf{x}-\mathbf{x}').$ 
Using this basis, we recover the usual definition of the recorded elastic wavefield in cartesian coordinates
\begin{equation}
  |\mathbf{u} \rangle  =  \sum_i \iiint^{+\infty}_{-\infty} u_i(\mathbf{x}) | i,  \mathbf{x} \rangle d^3x .
\end{equation}
(2) Next, we introduce the symbol  $| i , \mathbf{k} \rangle$ = $| \mathbf{\hat{e}}_i \rangle \otimes  | \mathbf{k} \rangle$ which denote a basis of eigenvectors of the vector Laplace operator 
 with eigenvalue $-\mathbf{k}^2$.  The symbol $ | \mathbf{k} \rangle$ represents
a  plane wave with wavenumber $\mathbf{k}$ which obeys the normalization of the continuum. By forming the tensor product 
$| \mathbf{\hat{e}}_i \rangle \otimes  | \mathbf{k} \rangle$, one obtains a plane wave perfectly polarized along direction 
$| \mathbf{\hat{e}}_i \rangle $ and propagating in direction $\mathbf{\hat{k}}$.
In the coordinate representation (1) introduced above, the eigenfunctions $| i , \mathbf{k} \rangle$  write 
 \begin{equation}
    \langle \mathbf{x},i   | j , \mathbf{k} \rangle =\dfrac{  \delta_{ij}  e^{i \mathbf{k}\cdot \mathbf{x}}}{(2 \pi)^{3/2}} .
 \end{equation}
(3) Finally, we denote by $ | \alpha ,  \mathbf{k} \rangle$ the eigenvectors of the elastodynamic operator
 in free space $L^0$, which is defined as
\begin{equation}
  \langle  \mathbf{x},i  | L^0 |\mathbf{u} \rangle = - (\lambda_0 + \mu_0)\partial_{x_i} \partial_{x_j} u_{j}(\boldsymbol{x})  
- \mu_0 \partial{x_j}\partial{x_j} u_{i}(\boldsymbol{x}).
\label{L0}
\end{equation}
 The notational conventions adopted in Eq. (\ref{L0}) are as follows:  
 the operator $L^0$ acts on the vector wavefunction $|\mathbf{u} \rangle  $ from the left to give
a new wavefunction $| \mathbf{v} \rangle = L^0 | \mathbf{u} \rangle  $, 
and the right-hand side gives the usual expression of the wavefunction $|\mathbf{v} \rangle$ in coordinate space.
The minus sign adopted in definition (\ref{L0}) makes $L^0$ a positive operator, i.e., such that $\langle u| L^0|u \rangle \geq 0$.
This property is easily proved since $\langle u| L^0|u \rangle /4$ represents the deformation energy of the solid 
(see \cite{benmenahem1998} p.31,  for details).  
 In coordinate space, the eigenvectors of $L^0$ are expressed as:
  \begin{equation} 
         \langle \mathbf{x} ,i     |  \alpha ,  \mathbf{k} \rangle
      =      \dfrac{\hat{p}_i^{\alpha} e^{i \mathbf{k} \cdot \mathbf{x}}}{(2 \pi)^{3/2}}
 \label{planewaves}
  \end{equation}
 where  $  \mathbf{\hat{p}}^{\alpha}$ denotes the polarization vector and 
$ \mathbf{k}$ is the wavevector. 
Equation (\ref{planewaves}) introduces the complete orthonormal set of  linearly polarized plane P and S waves. 
 In what follows, latin subscripts refer to cartesian coordinates (representations   1 and 2),  
while greek letters will 
be used to distinguish the contribution of the 3 different polarizations $\{S_1,S_2,P\}$ in representation 3.
 The  $  \mathbf{\hat{p}}^{\alpha}$ form a right-handed basis in polarization space
which coincides with the usual spherical coordinate frame $(\boldsymbol{\hat{\theta}},\boldsymbol{\hat{\phi}},\mathbf{\hat{r}}= \mathbf{\hat{k}})$,
where $\boldsymbol{\hat{\theta}}$ and $\boldsymbol{\hat{\phi}}$ point in the direction of increasing $\theta$ and $\phi$, respectively. Our conventions for the definition of the polarizations $\{S_1,S_2,P\}$ are depicted in Figure \ref{notation}.
It is important to keep in mind that polarization vectors rotate with the incident wavevector $ \mathbf{k}$. Note that with our
notational conventions, there should be no risk of confusion between representations (2) and (3). 

\subsection{Resolvent formula and Green's function retrieval}
 We now generalize the previous derivation to the case of an inhomogeneous elastic medium. We will consider simultaneously the case of 
a wavefield at  equipartition, or a system driven by a white noise distribution of random forces. Our formalism allows us to treat
 the two cases on the same footing. 
Some basic facts about resolvents and their connection with Green's functions are recalled hereafter.
Let us first introduce the operator $L$ which pertains to elastic waves in inhomogeneous media:
\begin{equation}
   L=L^0 + V(\omega_0^2),
\end{equation}
where $L^0$ is  the elastodynamic operator in free space introduced in equation (\ref{L0}) and
 $V(\omega_0^2)$ is the scattering potential at the frequency $\omega_0$:
\begin{equation}
    \langle \mathbf{x},i | V(\omega_0^2)   | l , \mathbf{x}'  \rangle =
            -\delta \rho(\mathbf{x}) \omega_0^2 \delta(\mathbf{x} - \mathbf{x}') \delta_{il} - 
          \partial_{x_j} \delta C_{ijkl}(\mathbf{x}) \partial_{x_k} \delta(\mathbf{x} - \mathbf{x}') ,
\label{scatpot}
\end{equation}
where  the usual summation convention  over repeated index is used. Equation (\ref{scatpot}) defines the matrix elements of the scattering potential operator in the coordinate representation. This representation is most usually encountered in the Green's function retrieval literature.
 For simplicity, we assume that the scattering potential  vanishes outside a bounded region of $\mathbb{R}^3$. 
To demonstrate that $V(\omega_0^2)$ is Hermitian, we
let $V(\omega_0^2)$ act on test functions $\mathbf{f}$ and $\mathbf{g}$. Since the Hermitian symmetry of the term involving the density perturbation 
$\delta \rho$ is easily verified, we only consider the $\delta C_{ijkl}$ term:
\begin{equation}
\begin{split}
   \langle \mathbf{f} | V(\omega_0^2) | \mathbf{g} \rangle= &
     \iiint_{\mathbb{R}^3}  f_i(\mathbf{x})^*  \partial_{x_j} \left(\delta C_{ijkl}(\mathbf{x}) \partial_{x_k} g_l(\mathbf{x})\right) d^3x   \\
    = &   - \iiint_{\mathbb{R}^3}    \left( \delta C_{ijkl}(\mathbf{x})  \partial_{x_j}  f_i(\mathbf{x}) \right)^* 
              \partial_{x_k} g_{l}(\mathbf{x}) d^3 x \\
       = & \iiint_{\mathbb{R}^3}  \left( \partial_{x_k}( \delta C_{ijkl}(\mathbf{x})  \partial_{x_j}  f_{i}(\mathbf{x}) \right)^*   g_{l}(\mathbf{x})  d^3 x \\
       = & \iiint_{\mathbb{R}^3}  \left( \partial_{x_k}( \delta C_{klij}(\mathbf{x})  \partial_{x_j}  f_{i}(\mathbf{x})) \right)^*   g_{l}(\mathbf{x})  d^3 x 
       =  \langle \mathbf{g} | V(\omega_0^2) | \mathbf{f} \rangle^*.
\label{vsymm}
\end{split}
\end{equation}
In the derivation of equation (\ref{vsymm}), we have used integration by parts twice and the usual symmetries of the elasticity tensor:
$C_{ijkl} = C_{klij}= C_{jikl} =C_{ijlk} $.  In the following, we assume that all Hermitian operators we have to deal with have unique  self-adjoint extensions.  Because the elastic tensor and  density fields are real, relation (\ref{vsymm}) 
implies that the matrix elements  of the scattering potential in  the representation (1) 
obey the following symmetry relation:
\begin{equation}
     \langle  \mathbf{x} ,i |V(\omega_0^2)  | j , \mathbf{x}'  \rangle  =
    V_{ij}(\mathbf{x},\mathbf{x}';\omega_0^2) =   V_{ji}(\mathbf{x}',\mathbf{x};\omega_0^2)
\label{symv}
\end{equation}

Let us introduce  the resolvent of a linear operator $L$ as follows: 
\begin{equation}
   (\lambda \mathcal{I} - L) R(\lambda) = \mathcal{I},
\label{resolv}
\end{equation}
where $\mathcal{I}$ is the identity operator in the space of vector wavefunctions, and  $\operatorname{Im} \lambda \neq 0$. Comparison of equations (\ref{resolv})
and (\ref{g0}) reveals that $G^0(\omega)$  is intimately related to $R^0(\lambda)$, the resolvent of $L^0$.
By making the following substitutions $\lambda \rightarrow \rho_0 (\omega + i/2\tau)^2$, $L \rightarrow L^0$, $R \rightarrow G^0(\omega)$
in equation (\ref{resolv}), we can formally define Green's function  at  angular frequency $\omega$ with a small, finite absorption time $\tau$.
 This justifies a posteriori the definition
of  Green's function adopted in section \ref{formalism}. With this choice,  $G^0(\omega)$ is related to the resolvent
of a \emph{positive} operator with continuous spectrum on the positive real axis (see Equation (\ref{L0}) and the discussion that follows).
Let us now introduce in  a similar fashion  the resolvent and  Green's function of the  equation of elastodynamics in an inhomogeneous medium as follows:
\begin{equation}
   \left[ \lambda \mathcal{I} -  \left( L^0 + V(\omega_0^2)\right) \right]R(\lambda) =  \mathcal{I},
\label{resolvinho}
\end{equation}
\begin{equation}
   \left[ \rho_0(\omega+i/2\tau)^2 \mathcal{I} -  \left( L^0 + V(\omega_0^2)\right) \right]G(\omega) =  \mathcal{I}.
\label{gfinho}
\end{equation}
An important point to be noted in the  definition (\ref{gfinho}) is the fact that $\omega$, the real part of the complex frequency, does not
necessarily equal $\omega_0$, the angular frequency at which the scattering potential is evaluated.
For elastic waves, the only physically accessible Green's function is obtained when we set $\omega=\omega_0$ in equation (\ref{gfinho}).
Taking the matrix elements of equation (\ref{gfinho}) in representation (1), we recover the fact that $G_{im}(\mathbf{x},\mathbf{x}';\omega_0)$ is the solution
of Equation (\ref{elasto1}) with the force density $f_{i} = -\delta_{im} \delta(\mathbf{x} -\mathbf{x}')$, as it should. 
 In equation (\ref{gfinho}), everything happens as if the
scattering potential has been ``frozen'' at the frequency $\omega_0$. This mathematical trick  allows us to use the standard tools
of potential scattering theory and facilitates intermediate calculations. In elastodynamics, this idea has previously been introduced by \cite{budreck1991}. 
Eventually, all key formulas will be obtained for $\omega=\omega_0$ and will involve physically accessible quantities only.

Let us consider  the elastic wavefield at frequency $\omega_1$ excited by a distribution of  forces and recorded at $\mathbf{x}_B$ in an inhomogeneous medium:
\begin{subequations}
   \begin{align}
   u_i(\mathbf{x}_B;\omega_1) = & -\sum_k \iiint_{-\infty}^{\infty}  \langle \mathbf{x}_B, i | G(\omega_1) | k, \mathbf{x}_1 \rangle
         \langle k, \mathbf{x}_1  |f \rangle d^3x_1,   \label{uixb1} \\
u_i(\mathbf{x}_B;\omega_1) = & - \sum_{\beta}\iiint^{\infty}_{-\infty}   \langle \mathbf{x}_B, i | G(\omega_1) |\beta,\mathbf{k}_1 \rangle \langle \beta,\mathbf{k}_1 |f \rangle d^3k_1,
  \label{uixb2}
\end{align}
\end{subequations}
In equation (\ref{uixb1}), we have used the coordinate representation to express the field generated by the force distribution. But any other complete
set of vectors can be inserted between $G$ and $f$.  As an example, in equation (\ref{uixb2}),  we make use of this freedom  by introducing a decomposition of the identity operator
 using the complete orthonormal set (3), which results in a hybrid representation. Indeed,  the symbol  $\langle \mathbf{x}_B, i | G(\omega_1) |\beta,\mathbf{k}_1 \rangle  =  G_{i \beta}(\mathbf{x}_B,\mathbf{k}_1;\omega_1)  $ has its left foot in representation (1) and its right foot in representation (3). The quantity $ G_{i \beta}(\mathbf{x}_B,\mathbf{k}_1;\omega_1)  $ 
can be calculated  from the usual coordinate representation
of Green's function $G_{ij}( \mathbf{x}_B, \mathbf{x}_1)$ by Fourier transformation over the variable $\mathbf{x}_1$ and by expanding the standard
 basis $| \mathbf{\hat{e}}_j \rangle$ over the rotating polarization basis $| \mathbf{\hat{p}}^{\beta} \rangle$. 
Clearly, the quantity $\tilde{f}_{\beta}(\mathbf{k}_1) = \langle \beta,\mathbf{k}_1 |f \rangle $ gives the amplitude of the plane wave with polarization
 $\beta$ and wavenumber $\mathbf{k}_1$, which is excited by the force distribution. As previously noted, if the force distribution is statistically homogeneous,
the wavenumber decomposition of the field has a formal character. Standard Fourier analysis can nevertheless be used after ensemble averaging (\cite{rytov1989t3}).
 Using equations (\ref{uixb1})-(\ref{uixb2}), the cross-correlation between the wavefields at $\mathbf{x}_A$ and $\mathbf{x}_B$ can be written as:
\begin{subequations}
\begin{multline}
 \llangle  u_i(\mathbf{x}_B;\omega_1)  u_j(\mathbf{x}_A;\omega_0)^* \rrangle = \\ \sum_{k,l} \iiint_{\mathbb{R}^6}  \langle \mathbf{x}_B, i | G(\omega_1) | k, \mathbf{x}_1 \rangle
    \langle \mathbf{x}_0, l | G(\omega_0)^{\dagger} | j, \mathbf{x}_A \rangle  (\llangle f_k(\mathbf{x}_1;\omega_1)  f_l(\mathbf{x}_0;\omega_0)^* \rrangle )   d^3x_0 d^3x_1
   \label{uxaxb1} 
\end{multline}
\begin{multline}
  \llangle  u_i(\mathbf{x}_B;\omega_1)  u_j(\mathbf{x}_A;\omega_0)^* \rrangle = \\ \sum_{\alpha,\beta} \iiint_{\mathbb{R}^6}  \langle \mathbf{x}_B, i | G(\omega_1) | \beta, \mathbf{k}_1 \rangle
    \langle \mathbf{k}_0, \alpha | G(\omega_0)^{\dagger} | j, \mathbf{x}_A \rangle   (\langle\!\langle \tilde{f}_{\beta}(\mathbf{k}_1;\omega_1)  \tilde{f}_{\alpha}(\mathbf{k}_0;\omega_0)^* \rrangle )   d^3k_0 d^3k_1 \label{uxaxb2},
\end{multline}
\end{subequations}
where $G(\omega_0)^{\dagger}$ denotes the Hermitian adjoint of $G(\omega_0)$. 
We now consider two possible physical assumptions regarding the excitations that drive the system. As in equation (\ref{randomforce1}), we may assume that the 
forces are randomly oriented and spatially delta-correlated. Alternatively, we may consider that the medium is illuminated by a  set of 
plane $P$ and $S$ waves whose  amplitudes constitute a white-noise  in modal space:
 \begin{equation}
     \llangle \tilde{f}_{\beta}(\mathbf{k}_1;\omega_1)  \tilde{f}_{\alpha}(\mathbf{k}_0;\omega_0)^* \rrangle = 2 \pi \delta_{\alpha \beta} \delta(\mathbf{k}_0 - \mathbf{k}_1) S(\omega_0) \delta(\omega_0 -\omega_1)
\label{equipartition}
\end{equation}
Assumption (\ref{equipartition}) may be interpreted as a requirement of equipartition: the system is illuminated by uncorrelated
 plane $P$ and $S$ waves coming from all possible directions with equal weights. This point will be further substantiated in section \ref{conclusion}.
Under which conditions such an equipartition state can be achieved in practice remains to be elucidated.
Reporting Equations (\ref{randomforce1}) and (\ref{equipartition}) in Equations (\ref{uxaxb1}) and (\ref{uxaxb2}) respectively, we find:
\begin{subequations}
\begin{align}
 C_{ij}(\mathbf{x}_B;\mathbf{x}_A;\omega_0)   = & S(\omega_0)  \sum_k \iiint^{\infty}_{-\infty}  
    \langle \mathbf{x}_B ,i | G(\omega_0) | k , \mathbf{x}  \rangle 
       \langle k , \mathbf{x} | G(\omega_0)^{\dagger} | j , \mathbf{x}_A  \rangle  d^3x  \\
  C_{ij}(\mathbf{x}_B;\mathbf{x}_A;\omega_0)     = &  S(\omega_0)   \sum_{\alpha} \iiint^{\infty}_{-\infty}  \langle \mathbf{x}_B, i | G(\omega_0) | \alpha, \mathbf{k}_0 \rangle
    \langle \mathbf{k}_0, \alpha | G(\omega_0)^{\dagger} | j, \mathbf{x}_A \rangle  d^3k_0  \\
    = &    S(\omega_0)   \langle \mathbf{x}_B,i |G(\omega_0) G(\omega_0)^{\dagger} |  j , \mathbf{x}_A  \rangle ,
  \label{resolventform}
\end{align}
\end{subequations}
where in the last equation, we have used the completeness
of the bases (1) and (3), respectively. In both cases -white noise distribution of forces, or wavefield at equipartition- we have reduced the computation of
 the cross-correlation of two random wavefields to the evaluation of the matrix elements of the operator 
$G(\omega_0) G(\omega_0)^{\dagger} $. In fact, from the defining property of  Green's function as the resolvent of a self-adjoint
 operator -equation (\ref{gfinho})-, the following identity follows:
\begin{equation}
\begin{split}
    G(\omega_0) G(\omega_0)^{\dagger} = & R(\rho_0 (\omega_0+i/2\tau)^2) R(\rho_0 (\omega_0 -i/2\tau)^2)   \\
                                     = & -\frac{ i \tau}{\rho_0 \omega_0} \left( R(\rho_0 (\omega_0+i/2\tau)^2) -  R(\rho_0 (\omega_0-i/2\tau)^2)  \right)\\
                                     = & -\frac{ \tau}{2 i \rho_0 \omega_0} (G(\omega_0) - G(\omega_0)^{\dagger}),
\label{g0rab}
\end{split}
\end{equation}
with $R(\lambda) $ the resolvent operator introduced in Equation (\ref{resolvinho}).
The key point in the above derivation is the use of a fundamental operator equation known
as the ``first resolvent formula'' in the second line of Equation (\ref{g0rab}). The resolvent formula is 
valid for a broad class of operators (closed linear operators) including the differential operators considered in this work \citep{richtmyer1978}.
 The last step follows simply from the
Hermitian character of the elastodynamic operator $L^0 + V(\omega_0^2)$. 
Taking into account the power spectral density of the source and calculating the matrix elements of the right-hand side of equation (\ref{g0rab}) yields:
\begin{multline}
\frac{d}{dt} \lim_{T \to \infty } \frac{1}
 {T}\int_{ - T/2}^{T/2}  \llangle u_i\left( {{\boldsymbol{x}}_B ,t'} \right) {u_j\left( {{\boldsymbol{x}}_A ,t' - t } \right)} \rrangle dt' 
  = \\
  \frac{\tau }{2 \rho_0  (2 \pi) }  \int_{ - \infty }^\infty  {d\omega } \left[ {e^{ - i\omega t }
  G_{ij}\left( {{\boldsymbol{x}}_B ,{\boldsymbol{x}}_A,\omega  } \right) - e^{ - i\omega \left( { - t } \right)}  
 G_{ji}\left( {\boldsymbol{x}}_A,{\boldsymbol{x}}_B ,\omega   \right)} \right] S\left( \omega  \right),
 \label{gfra2}
 \end{multline}
where the Hermitian symmetry of $G(\omega)$ has been used. When the reciprocity theorem applies, the integrand in the right-hand side of equation (\ref{gfra2}) is 
recognized as the usual imaginary part of  Green's function. Thus, the derivative of the cross-correlation tensor of random signals yields a filtered
version of the difference between advanced and retarded Green's functions.
 Since $G$ can be substituted with the Green's function of any self-adjoint operator, this
implies the reconstruction of  Green's function for other systems governed by linear wave equations.
Derivations of Green's function reconstruction based on a different operator formalism  have been further developped
by \cite{Wapenaar2006} and \cite{sniederwapenaar2007}. Our derivation highlights the fact that a wavefield at equipartition 
and a random distribution of white noise sources are indistinguishable from the point of view of
cross-correlation functions.  The complete equivalence between the two types of excitations will be demonstrated in section \ref{conclusion}.
The general result (\ref{g0rab}) does not constitute, however,
the end of our investigations. We will show that interesting information on the scattering of elastic waves can
be obtained by expressing  the Green's function of an inhomogeneous medium using the formalism of scattering theory.
To do so, basic  notions of scattering of elastic waves are recalled in the next section.
\section{Summary of scattering of elastic waves}
\label{scattering}
Our starting point is the observation that  Equation (\ref{gfinho}) is formally identical to a Schr\"odinger equation, with
a scattering potential $V(\omega_0^2)$. This allows us to use all the arsenal of  scattering theory.
In particular we can introduce the retarded Lippman-Schwinger eigenvectors  of the operator $(L^0+V(\omega_0^2))$
 which satisfy the following equation in the infinitesimal absorption limit $\tau \to \infty$ (\cite{Economou.2006}):
\begin{equation}
| \psi^{\alpha}(k \mathbf{\hat{k}}) \rangle = |\alpha,k \mathbf{\hat{k}} \rangle + G^0(c_{\alpha}k)V(\omega_0^2) 
| \psi^{\alpha}(k \mathbf{\hat{k}}) \rangle,
\label{lippman}
\end{equation}
with $k>0$. 
It is worth noting that in equation (\ref{lippman}), the potential is ``frozen'' at the frequency $\omega_0$,
and is scanned by plane $P$ and $S$ waves with all possible wavevectors $k \mathbf{\hat{k}}$ with $k>0$, and eigenfrequency
 $\omega= c_{\alpha} k >0 $. These eigenvectors do not depend on the sign of $\omega_0$. 
For elastic waves, the physical  wavefunction is obtained when the condition $\omega=\omega_0=c_{\alpha} k$ is satisfied.
For later developments it is nevertheless convenient to introduce the more general Lippmann-Schwinger eigenvectors (\ref{lippman}).
All the scattering properties of an inhomogeneous elastic medium are conveniently
encapsulated in an operator denoted by $T^{(\omega_0^2)}$ which is defined as follows:
 \begin{equation}
  T^{(\omega_0^2)}(\omega) = V(\omega_0^2) + V(\omega_0^2)G(\omega)V(\omega_0^2)  .
\label{tmatop}
\end{equation} 
In equation (\ref{tmatop}), we have carefully distinguished between the angular frequency of the target -$\omega_0$- and of 
Green's function -$\omega$-. Clearly, the physical result is obtained when $\omega$ equals $\omega_0$ and we introduce a special notation
for this case:
 \begin{equation}
    T^0 = T^{(\omega_0^2)}(\omega_0)
\end{equation} 
For simplicity, we have dropped the frequency dependence in this definition. The superscript $^0$ reminds us of the
fact that  $\omega=\omega_0$.
 The definition of the $T^{(\omega_0^2)}$-operator
 adopted in equation (\ref{tmatop}) is in keeping with our definition of  Green's function of an inhomogeneous medium  (see equation (\ref{gfinho}))
 and will facilitate intermediate calculations in section \ref{formalproof}. 
Using the $T^{(\omega_0^2)}$-operator, we can rewrite the Lippman-Schwinger equation as follows  ($\tau \to \infty$, $k>0$): 
\begin{equation}
  | \psi^{\alpha}(k \mathbf{\hat{k}}) \rangle = |\alpha,k \mathbf{\hat{k}} \rangle + 
G^0(c_{\alpha}k)T^{(\omega_0^2)}(c_{\alpha}k) |\alpha,k \mathbf{\hat{k}} \rangle ,
\label{lippman2}
\end{equation}
which further implies:
\begin{equation}
 V(\omega_0^2)  | \psi^{\alpha}(k \mathbf{\hat{k}}) \rangle  = T^{(\omega_0^2)}(c_{\alpha}k) |\alpha,k \mathbf{\hat{k}} \rangle 
\label{vpsitphi}
\end{equation} 
Green's function of the heterogeneous medium can also  conveniently be rewritten in terms of the $T^{(\omega_0^2)}$-operator
and the free space Green's function alone:
\begin{equation}
  G(\omega) = G^0(\omega) + G^0(\omega)T^{(\omega_0^2)}(\omega)G^0(\omega).
\label{g0tg0}
\end{equation}

Since our approach is based on an analogy with quantum scattering, a word of comment about the sign of $\omega_0$ is in order at this point.
The problem of negative frequencies is easily solved because  we  consider real fields in the space-time domain. This imposes the following Hermitian symmetry
 on the matrix elements of $G$ and $T$ in coordinate representation:
 \begin{align}
      G_{ij}(\mathbf{x},\mathbf{x'};-\omega_0) = & G_{ij}(\mathbf{x},\mathbf{x'};\omega_0)^* &    
   T^{(\omega_0^2)}_{ij}(\mathbf{x},\mathbf{x'};-\omega_0) = & T^{(\omega_0^2)}_{ij}(\mathbf{x},\mathbf{x'};\omega_0)^*
 \label{omsymm}
\end{align}
As a consequence, all results obtained for positive frequency are readily translated to negative frequency using the symmetry conditions (\ref{omsymm}). 

In the far-field of the scatterer, i.e. in the limit $x \gg D$ with $\mathbf{x}$ the observation point and $D$ the dimension of the scatterer
 (Figure \ref{pbsetting}), we may approximate  the free space Green's function as follows:
\begin{equation}
 G^0_{ij}(\mbf{x},\mbf{x}') \approx - \frac{e^{i k_s x -i k_s \mbf{\hat{x}} \cdot \mbf{x}'}  }{4 \pi \rho_0 c_s^2 x }   \left( \delta_{ij} - \hat{x}_i \hat{x}_j \right) 
                           - \frac{e^{i k_p x -i k_p \mbf{\hat{x}} \cdot \mbf{x}'}}{4 \pi \rho_0 c_p^2 x} \hat{x}_i \hat{x}_j,
\label{gasym}
 \end{equation}
where $\mathbf{x'}$ denotes an arbitrary point of the scatterer. The position vectors are measured from an arbitrary origin located inside the scattering region
 and $\mathbf{\hat{x}}$ is a unit vector in the direction of $\mathbf{x}$.
The asymptotic formula (\ref{gasym}) allows us to express the Lippman-Schwinger wavefunction 
in the usual coordinate representation as follows:
\begin{equation}
  \langle i , \mathbf{x} |  \psi^{\beta}(k^{\beta} \mathbf{\hat{k}}_b) \rangle=
 \dfrac{1}{(2\pi)^{3/2}} \left(  \hat{q}^{\beta}_i e^{i k^{\beta} \mathbf{\hat{k}}_b \cdot \mathbf{x}} + 
  \dfrac{ f^{\alpha \leftarrow \beta}(\mathbf{\hat{k}}_a,\mathbf{\hat{k}}_b)}{x} 
   \hat{p}^{\alpha}_i e^{i k^{\alpha} x} \right),
\label{asymlipp}
\end{equation}
valid in the limit $x \to \infty$, where $x=|\mathbf{x}|$.
In equation (\ref{asymlipp}), we have introduced the scattering amplitudes:
\begin{equation}
\begin{split}
 f^{\alpha \leftarrow \beta}(\mathbf{\hat{k}}_a,\mathbf{\hat{k}}_b)  = & -\frac{(2 \pi)^3}{4\pi \rho_0 (c_{\alpha})^2}  
 \langle  \alpha , k^{\alpha} \mathbf{\hat{k}}_a | T^{(\omega_0^2)}(\omega_0) |  \beta , k^{\beta} \mathbf{\hat{k}}_b  \rangle \\
                                               = & -\frac{(2 \pi)^3}{4\pi \rho_0 (c_{\alpha})^2}  
                                T^0_{\alpha \beta} ( k^{\alpha} \mathbf{\hat{k}}_a  , k^{\beta} \mathbf{\hat{k}}_b)
\label{scatamp}
\end{split}
\end{equation}
and the notations: $k^{\alpha,\beta} = \omega_0/c_{\alpha,\beta} >0$.
In the last line of equation (\ref{scatamp}), we have introduced
``on-shell'' matrix elements of the $T^0$-operator. 
In general, the matrix elements   $\langle \alpha,\mathbf{k}_a| T^{(\omega_0^2)}(\omega)| \beta, \mathbf{k}_b \rangle$ are said to be ``on-shell'' when the
condition $\omega^2 = c^2_{\alpha}k^2_a = c^2_{\beta}k^2_b $ is satisfied. 
Although this property does not depend on the fact that we are on the shell $\omega=\omega_0$, we eventually  specialize our results
to this particular case. 
The scattering amplitudes contain information on the scattering
pattern for all possible mode conversions. For instance, taking $\beta=1$, $\alpha=3$, corresponds to the physical situation
where a linearly polarized incident shear wave is scattered into a compressional wave.  Note that the definition is valid
for an arbitrary anisotropic inhomogeneity.  From the symmetries of the scattering potential: 
$V_{ij}(\mathbf{x},\mathbf{x}';\omega_0^2) = V_{ji}(\mathbf{x}',\mathbf{x};\omega_0^2)$ and  the reciprocity of the free space Green's function 
$G^0_{ij}(\mathbf{x},\mathbf{x}';\omega)= G^0_{ji}(\mathbf{x}',\mathbf{x};\omega)$, we deduce a similar reciprocity relation
for the $T^{(\omega_0^2)}$ operator:
\begin{equation}
 T^{(\omega_0^2)}_{ij}(\mathbf{x},\mathbf{x}';\omega) = T^{(\omega_0^2)}_{ji}(\mathbf{x}',\mathbf{x};\omega).
\label{reciprotop}
\end{equation}
Using equation (\ref{reciprotop}), our formalism allows us to recover well-known reciprocity relations for
the scattering amplitudes:
 \begin{equation}
\begin{split}
 T^{(\omega_0^2)}_{\alpha \beta}(\mathbf{k}_a,\mathbf{k}_b;\omega) = & \sum_{i,j} \dfrac{1}{(2 \pi)^3}
 \iiint_{\mathbb{R}^6} \hat{p}^{\alpha}_i e^{-i \mathbf{k}_a \cdot \mathbf{x}} T^{(\omega_0^2)}_{ij}(\mathbf{x},\mathbf{x}';\omega) 
      \hat{q}^{\beta}_j  e^{i \mathbf{k}_b \cdot \mathbf{x}'} d^3xd^3x'  \\
            = & \sum_{i,j} \dfrac{1}{(2 \pi)^3}
 \iiint_{\mathbb{R}^6} \hat{q}^{\beta}_j e^{i \mathbf{k}_b \cdot \mathbf{x}'} T^{(\omega_0^2)}_{ji}(\mathbf{x}',\mathbf{x};\omega) 
      \hat{p}^{\alpha}_i  e^{-i \mathbf{k}_a \cdot \mathbf{x}} d^3xd^3x' \\
          = & \text{sgn}(\alpha,\beta)  T^{(\omega_0^2)}_{\beta \alpha}(-\mathbf{k}_b,-\mathbf{k}_a;\omega),
\end{split}
\label{symtop}
 \end{equation}
where $\text{sgn}(\alpha,\beta)= -1$ for $(\alpha,\beta) \in \{(1,2),(2,1),(1,3),(3,1) \}$ and $+1$ otherwise.
 In equation (\ref{symtop}), the matrix elements are generally off-shell because the condition $\omega^2 = c^2_{\alpha} k^2_a=c^2_{\beta} k^2_b$ needs
not be satisfied. This relation is in fact far more general than what is required for our purposes. We will therefore specialize it to the
on-shell case with the condition $\omega=\omega_0$.
 Note that the sign change -$\text{sgn}(\alpha,\beta)$- is a mere consequence of the convention adopted to define the polarization vectors
and illustrated in Figure \ref{notation}. Indeed, upon a change
of the wavevector $\mathbf{k}_a \rightarrow -\mathbf{k}_a$, the polarization vectors transform as follows: 
$(\mathbf{\hat{p}}^1,\mathbf{\hat{p}}^2,\mathbf{\hat{p}}^3) \rightarrow (\mathbf{\hat{p}}^1,-\mathbf{\hat{p}}^2,-\mathbf{\hat{p}}^3) $.
As an example, let us consider an incident $S$ wave with polarization vector $\mathbf{\hat{q}}^{1}$ and a scattered $P$ wave
with polarization vector  $\mathbf{\hat{p}}^{3}$. Application of the symmetry relation (\ref{symtop}) to equation (\ref{scatamp}) yields the following
reciprocity relation between the scattering amplitudes:
\begin{equation}
 c_p^2  f^{P \leftarrow S_1 } (\mathbf{\hat{k}}_a,\mathbf{\hat{k}}_b)  =  -c_s^2 f^{S_1 \leftarrow P } (-\mathbf{\hat{k}}_b,-\mathbf{\hat{k}}_a),
\label{reciprocityf}
\end{equation}
a relation which has been noted by a number of authors in the past \citep{varatharajulu1977,dassios1987,aki1992}.
Interestingly, this reciprocity relation shows up without any effort from the formalism we have introduced. Some physical
implications of relation (\ref{reciprocityf}) will be discussed in section \ref{conclusion},  in connection with the role
of equipartition in Green's function retrieval.
With the help of our scattering formalism, we will explore the consequences of  Green's function reconstruction
in a scattering medium in the next section.

\section{Green's function reconstruction in a scattering medium }
\label{general}
As noted in section (\ref{formalism}),  Green's function reconstruction from homogeneous random sources is a consequence of a basic
operator identity. We now explore the consequences of this result in the case of a scattering medium.  
The  identity satisfied by  Green's function of the heterogeneous medium writes:
 \begin{equation}
 G(\omega_0) G(\omega_0)^{\dagger}  =  -\frac{ \tau}{2 i \rho_0 \omega_0} (G(\omega_0) - G(\omega_0)^{\dagger}).
\label{gfrab}
 \end{equation}
Using equations (\ref{g0tg0})  we can express the left-hand side (L.H.S.)  of equation (\ref{gfrab}) as follows:
\begin{align}
 G(\omega_0) G(\omega_0)^{\dagger} = & G^0(\omega_0) G^0(\omega_0)^{\dagger} + 
                 G^0(\omega_0)G^0(\omega_0)^{\dagger}T^{0 \dagger}G^0(\omega_0)^{\dagger} \\
                 &   + G^0(\omega_0) T^0 G^0(\omega_0)G^0(\omega_0)^{\dagger}  \\
            &   +   G^0(\omega_0)T^0 G^0(\omega_0)G^0(\omega_0)^{\dagger}T^{0 \dagger}G^0(\omega_0)^{\dagger} . \label{ggdag}
\end{align}
Similarly, we obtain for the right-hand side (R.H.S) of equation (\ref{gfrab}):
\begin{equation}
    G(\omega_0) - G(\omega_0)^{\dagger}            =   G^0(\omega_0) -  G^0(\omega_0)^{\dagger} + G^0(\omega_0) T^0  G^0(\omega_0) 
                                                 -   G^0(\omega_0)^{\dagger} T^{0 \dagger}  G^0(\omega_0)^{\dagger}           .
\end{equation}
Upon inserting relation (\ref{g0rab}) into equation (\ref{ggdag}),  Green's function reconstruction formula (\ref{gfrab}) 
 yields the following identity for the $T$ operator:
\begin{equation}
      \frac{ 1}{2 i } \left(T^0 - T^{0 \dagger}\right) =
  -\frac{\rho_0 \omega_0}{\tau}  T^0 G^0(\omega_0) G^0(\omega_0)^{\dagger} T^{0 \dagger} 
\label{optthab}
\end{equation} 
In order to elucidate the physical meaning of equation (\ref{optthab}), let us take matrix elements in the representation (3).
We find: 
\begin{equation}
     \frac{ 1}{2 i }   \langle \alpha , \mathbf{k}_a | \left(T^0 - T^{0 \dagger}\right) | \beta ,\mathbf{k}_b \rangle =
       \frac{ 1}{2 i } \left( T^0_{\alpha \beta}( \mathbf{k}_a,\mathbf{k}_b) -  T^0_{\beta \alpha}( \mathbf{k}_b,\mathbf{k}_a)^*      \right),
\label{optth1}
\end{equation}
on the L.H.S., and:
\begin{multline}
 -\frac{\rho_0 \omega_0}{\tau}  \langle \alpha , \mathbf{k}_a |  T^0 G^0(\omega_0) G^0(\omega_0)^{\dagger} T^{0 \dagger} | \beta ,\mathbf{k}_b \rangle  = \\
- \dfrac{\omega_0}{\rho_0 \tau} \sum_{\gamma} \iiint_{\mathbb{R}^3} 
 \dfrac{ \langle {\alpha} , \mathbf{k}_a |  T^0 | \gamma, \mathbf{k} \rangle \langle \gamma, \mathbf{k}|  T^{0 \dagger} 
| {\beta} , \mathbf{k}_b \rangle   }{\left(\omega_0^2 - c_{\gamma}^2k^2 \right)^2 + \omega_0^2/\tau^2}  d^3k ,
\label{previouseq}
\end{multline}
on the R.H.S., where we have made use of the spectral decomposition of $G^0(\omega_0)$, valid irrespective of the sign of $\omega_0$:
\begin{equation}
 G^0(\omega_0) = \sum_{\gamma}\iiint_{\mathbb{R}^3}  \dfrac{|\gamma, \mathbf{k} \rangle \langle\gamma,\mathbf{k} |}{\rho_0(\omega_0^2 -c_{\gamma}^2k^2 + i \omega_0/\tau)}d^3k
\label{g0spec}
\end{equation}
 On the right-hand side of equation (\ref{g0spec}), the symbol $|\gamma, \mathbf{k} \rangle \langle\gamma,\mathbf{k} |$ is
an operator which projects the wavefunctions on the subspace of plane $P$ and $S$ waves with wavevector $\mathbf{k}$.  
We will now turn equation (\ref{previouseq}) into a more physically appealing form. First, we note that the representation
of the product of Green's function $G^0(\omega_0)G^0(\omega_0)^{\dagger}$ is left unchanged by the substitution $\mathbf{k} \rightarrow -\mathbf{k}$. 
Next, we  make use of the reciprocity relation (\ref{symtop}) to rearrange the numerator as follows:
\begin{equation}
  \langle {\alpha} ,  \mathbf{k}_a |  T^0| \gamma, -\mathbf{k} \rangle
  \langle \gamma, -\mathbf{k}|  T^{0 \dagger} | {\beta} , \mathbf{k}_b  \rangle =  
   \text{sgn}(\alpha,\beta)   \langle   {\beta} ,-  \mathbf{k}_b  |  T^{0 \dagger} |  \gamma, \mathbf{k}   \rangle
   \langle \gamma, \mathbf{k}  |  T^0 |  {\alpha} ,- \mathbf{k}_a \rangle,
\label{optth3}
\end{equation}
where we have used $\text{sgn}(\alpha,\gamma) \text{sgn}(\gamma,\beta)  = \text{sgn}(\alpha,\beta) $.
The following substitutions: $ \mathbf{k}_a \rightarrow - \mathbf{k}_b $, 
$  \mathbf{k}_b \rightarrow -  \mathbf{k}_a $, $\alpha \rightarrow \beta$, $\beta \rightarrow \alpha$, leaves  equation (\ref{optth1}) unchanged except for an additional  factor $\text{sgn}(\alpha,\beta)$ which cancels the same factor on the right-hand side
of equation (\ref{optth3}).   We therefore arrive at the following general identity:
\begin{equation}
  \dfrac{1}{2i} \left( T^0_{\alpha \beta}( \mathbf{k}_a,  \mathbf{k}_b) -
     T^0_{\beta \alpha}(\mathbf{k}_b, \mathbf{k}_a)^* \right) = 
  -\dfrac{\omega_0}{\rho_0 \tau} \sum_{\gamma} \iiint  \dfrac{ T^0_{\gamma \beta}(\mathbf{k}, \mathbf{k}_b  ) 
       T^0_{\gamma \alpha}( \mathbf{k},  \mathbf{k}_a  )^* }
       {  \left( \omega_0^2 - c_{\gamma}^2k^2\right)^2 + \omega_0^2/\tau^2   }    d^3k   ,
\label{genoptthabs}
\end{equation}  
which we will call the generalized optical theorem in the presence of small finite absorption.
Equation (\ref{genoptthabs}) is a general identity which must be obeyed by the ``off-shell''  matrix elements of 
 the $T^0$-operator for a general scatterer in an elastic medium. The matrix elements are said to be \emph{off-shell}
because  $\mathbf{k}$, $\mathbf{k}_a$ and $\mathbf{k}_b$ do not have to verify the shell condition $\omega^2_0 =c^2_{\gamma} k^2 $.  The physical meaning of these off-shell contributions will be discussed below.
 Equation (\ref{genoptthabs}) is a central result of this paper, 
from which all forms of optical theorem derived in the literature can be deduced.
The justification of the term ``generalized optical theorem'' is made clear when we consider the limit of
infinitesimal absorption $\tau \to \infty$. Using standard results from distribution theory, we find:
  \begin{equation}
  \lim_{\tau \to \infty} - \dfrac{\omega_0/(\rho_0 \tau)}{\left( \omega_0^2 - c_{\gamma}^2k^2\right)^2 + \omega_0^2/\tau^2 } = 
       - \dfrac{\pi}{2 \rho_0 \omega_0 c_{\gamma}} \left[ \delta \left( \dfrac{\omega_0}{c_{\gamma}} -k  \right)  +  
      \delta \left(\dfrac{\omega_0}{c_{\gamma}}+k \right) \right]
\label{funclim}
    \end{equation}
Upon inserting the previous relation in equation (\ref{genoptthabs}) and integrating over the wave number $k$, we arrive at:
\begin{equation}
\dfrac{1}{2i} \left(  T^0_{\alpha \beta}(\mathbf{k}_a,\mathbf{k}_b) - T^0_{\beta \alpha}(\mathbf{k}_b,\mathbf{k}_a)^*   \right) = 
-\dfrac{\pi \omega_0}{2 \rho_0}   \sum_{\gamma}\dfrac{1}{(c_{\gamma})^3} \iint_{4 \pi}  
 T^0_{\gamma \beta} \left( \dfrac{\omega_0}{c_{\gamma}} \mathbf{\hat{k}}, \mathbf{k}_b  \right)        
T^0_{\gamma \alpha} \left(\dfrac{\omega_0}{c_{\gamma}}  \mathbf{\hat{k}},  \mathbf{k}_a  \right)^* d^2\hat{k}   ,
\label{genoptthnoabs}
\end{equation} 
which is the form of the generalized optical theorem in a medium with infinitesimal absorption.
In Equation (\ref{genoptthnoabs}), the symbol  $d^2\hat{k}$ denotes the element of solid angle in wavenumber space.
 Again, off-shell
terms of the $T$ matrix appear on both side of equation (\ref{genoptthnoabs}). On the right hand side, two arguments
are on the  shell $c^2_{\gamma} k^2  = \omega^2_0 $. Physically, off-shell terms are related to the
physical situation where source and detection take place in the near field of the scatterer \citep{peierls1979}.
Thus the reconstruction of  Green's function for arbitrary locations of the receivers in the medium imposes
symmetry relations that pertain to the off-shell $T$-matrix.
More examples of the optical theorem for electromagnetic and quantum waves can be found in the book of \cite{newton2002}.
 If we now impose $\alpha=\beta$ and further restrict all wavenumbers
to the shell $\omega_0$, we arrive at classical forms of the optical theorem:
\begin{equation}
 \text{Im} T^0_{\alpha \alpha}\left(\dfrac{\omega_0}{c_{\alpha}}\mathbf{\hat{k}}_a , \dfrac{\omega_0}{c_{\alpha}} \mathbf{\hat{k}}_a \right) = 
       -\dfrac{\pi \omega_0}{2 \rho_0} \sum_{\gamma} \dfrac{1}{c_{\gamma}^3} \iint\limits_{4 \pi} 
 \left| T^0_{\gamma \alpha}\left(\dfrac{\omega_0}{c_{\gamma}}  \mathbf{\hat{k}},\dfrac{\omega_0}{c_{\alpha}}  \mathbf{\hat{k}}_a \right) \right|^2 d^2\hat{k}
\label{clasoptth}
\end{equation} 
 As discussed in \cite{peierls1979}, the on-shell matrix elements describe the angular dependence of the scattered waves in 
the far-field of the heterogeneity.
The physical meaning of equation (\ref{clasoptth}) for elastic waves will be further elucidated in section \ref{formalproof}.
It is important to keep in mind that all the relations obtained so far have been based on  the Green's function reconstruction 
theorem. In the next section, we make the link with the scattering theory developed in section \ref{scattering}. 
\section{Generalized optical theorem: formal derivations}
\label{formalproof}
In this section, we provide  derivations of the various forms of optical theorems obtained within the framework of Green's function reconstruction.
 First, we note that Equation (\ref{clasoptth}) which is the on-shell version of equation (\ref{genoptthnoabs}), can be obtained straightforwardly from the scattering 
theory developed in section (\ref{scattering}):
\begin{multline}
\dfrac{1}{2i} \left[  T^0_{\alpha \beta}\left(\dfrac{\omega_0}{c_{\alpha}} \mathbf{\mathbf{\hat{k}}}_a,\dfrac{\omega_0}{c_{\beta}}\mathbf{\hat{k}}_b \right) -
 T^0_{\beta \alpha} \left(\dfrac{\omega_0}{c_{\beta}}\mathbf{\hat{k}}_b,\dfrac{\omega_0}{c_{\alpha}}\mathbf{\hat{k}}_a \right)^*   \right]
=  \\
   \dfrac{1}{2i}  \left[ \langle \alpha , \dfrac{\omega_0}{c_{\alpha}} \mathbf{\hat{k}}_a |  T^0
      | \beta , \dfrac{\omega_0}{c_{\beta}}\mathbf{\hat{k}}_b  \rangle 
    -  \langle \alpha ,  \dfrac{\omega_0}{c_{\alpha}} \mathbf{\hat{k}}_a |  T^{0 \dagger} | 
        {\beta} , \dfrac{\omega_0}{c_{\beta}} \mathbf{\hat{k}}_b  \rangle \right]  \\
 =  \dfrac{1}{2i} \left[   \langle {\alpha} ,  \dfrac{\omega_0}{c_{\alpha}} \mathbf{\hat{k}}_a |  V(\omega_0^2)  |
     \psi^{\beta} \left(\dfrac{\omega_0}{c_{\beta}} \mathbf{\hat{k}}_b \right)  \rangle 
           -  \langle  \psi^{\alpha} \left( \dfrac{\omega_0}{c_{\alpha}} \mathbf{\hat{k}}_a\right) |  V(\omega_0^2)  |
         \beta , \dfrac{\omega_0}{c_{\beta}} \mathbf{\hat{k}}_b \rangle   \right] \\
    =  \dfrac{1}{2i} \left[ \langle  \psi^{\alpha} \left( \dfrac{\omega_0}{c_{\alpha}} \mathbf{\hat{k}}_a \right) |  V(\omega_0^2)  | \psi^{\beta} \left(\dfrac{\omega_0}{c_{\beta}} \mathbf{\hat{k}}_b \right)  \rangle   -   \langle  \psi^{\alpha} \left( \dfrac{\omega_0}{c_{\alpha}} k^{\alpha} \mathbf{\hat{k}}_a \right) |  V(\omega_0^2) G^0(\omega_0)^{\dagger}  V(\omega_0^2) |
   \beta , \dfrac{\omega_0}{c_{\beta}} \mathbf{\hat{k}}_b \rangle   \right] \\
       -    \dfrac{1}{2i} \left[ \langle  \psi^{\alpha} \left( \dfrac{\omega_0}{c_{\alpha}}  \mathbf{\hat{k}}_a\right) |  V(\omega_0^2)  |
 \psi^{\beta} \left(\dfrac{\omega_0}{c_{\beta}} \mathbf{\hat{k}}_b \right)  \rangle 
        +    \langle  \psi^{\alpha} \left( \dfrac{\omega_0}{c_{\alpha}}  \mathbf{\hat{k}}_a \right) |  V(\omega_0^2) G^0(\omega^0)  V(\omega_0^2)  | 
         \psi^{\beta} \left( \dfrac{\omega_0}{c_{\beta}} \mathbf{\hat{k}}_b \right)  \rangle   \right]  \\
   =   \dfrac{1}{2i} \left[ \langle  \psi^{\alpha} \left( \dfrac{\omega_0}{c_{\alpha}}  \mathbf{\hat{k}}_a \right) | 
     V(\omega_0^2) \left( G^0(\omega^0) - G^0(\omega^0)^{\dagger}\right)
        V(\omega_0^2)  | \psi^{\beta} \left(\dfrac{\omega_0}{c_{\beta}} \mathbf{\hat{k}}_b \right)  \rangle   \right]  \\
  =   \dfrac{1}{2i} \left[  \langle {\alpha} ,  \dfrac{\omega_0}{c_{\alpha}}  \mathbf{\hat{k}}_a |  T^{0 \dagger}   \left( G^0(\omega^0) - G^0(\omega^0)^{\dagger} \right)   
    T^0 | {\beta} ,\dfrac{\omega_0}{c_{\beta}} \mathbf{\hat{k}}_b  \rangle  \right]   \\
  =  -\dfrac{\pi \omega_0}{2 \rho_0}   \sum_{\gamma}\dfrac{1}{(c_{\gamma})^3} \iint_{4 \pi}  
 T^0_{\gamma \beta} \left( \dfrac{\omega_0}{c_{\gamma}} \mathbf{\hat{k}}, \dfrac{\omega_0}{c_{\beta}} \mathbf{\hat{k}}_b  \right)        
T^0_{\gamma \alpha} \left(\dfrac{\omega_0}{c_{\gamma}}  \mathbf{\hat{k}},   \dfrac{\omega_0}{c_{\alpha}}  \mathbf{\hat{k}}_a  \right)^* d^2\hat{k} 
\label{onshgenoptthnoabs}
\end{multline}
where the infinitesimal absorption limit ($\tau \to \infty$) is understood throughout. The first equality follows from the definition of the $T^0$-operator. 
In the second equality,
we make use of the relation (\ref{vpsitphi}) between Lippman-Schwinger eigenvectors and plane waves. The third equality follows from the Lippman-Schwinger
equation (\ref{lippman2}). The fifth equality makes again use of (\ref{vpsitphi}).  We recover the generalized optical theorem (\ref{genoptthnoabs}) after
inserting the spectral decomposition of the free space Green's function (\ref{g0spec}) in the sixth equality.   Some on-shell forms of the optical theorem similar to
 (\ref{clasoptth})-(\ref{onshgenoptthnoabs}) have been previously
derived  in \cite{lu2011} for a homogeneous and isotropic heterogeneity, and \cite{budreck1992} for a general heterogeneity.
 These formulas are less general than the fully off-shell 
expressions (\ref{genoptthabs}) and (\ref{genoptthnoabs}).
The physical meaning of the formula (\ref{clasoptth}) can be further elucidated by introducing the scattering amplitudes
defined in (\ref{scatamp}). From the definition of the scattering cross-section   as the ratio between the total 
 energy scattered per unit time, normalized by the energy flux density of the incoming plane wave:
\begin{equation}
 \sigma_{\alpha}(\mathbf{\hat{k}}_a) = \sum\limits_{\beta} \dfrac{ \rho_0 \omega^2_0 c_{\beta} }{ \rho_0 \omega_0^2 c_{\alpha}}
 \iint_{4 \pi} |   f^{\beta \leftarrow \alpha}(\mathbf{\hat{k}}_b,\mathbf{\hat{k}}_a)  |^2 d^2 \hat{k}_b,
\end{equation}
we arrive at the following simple formula:
 \begin{equation}
 \sigma_{\alpha}(\mathbf{\hat{k}}_a) = \frac{4 \pi}{k_{\alpha}} \text{Im}f^{\alpha \leftarrow \alpha} (\mathbf{\hat{k}}_a,\mathbf{\hat{k}}_a),
\label{clasoptth2}
\end{equation}
which has been derived by other authors in the past (\cite{varatharajulu1977,dassios1987}). The physical meaning of equation (\ref{clasoptth})
is now clear: it expresses the conservation of energy in the scattering process. Energy conservation is realized thanks to the interference between the incident wave and the wave scattered exactly in the  forward direction in the far-field.

To prove more general versions of the optical theorem, we invoke the spectral decomposition of Green's
function of the heterogeneous medium in terms of the Lippman-Schwinger eigenvectors. We make no attempt to justify rigorously
this decomposition and proceed completely formally. Mathematical foundations of this approach  can be found
in the book of \cite{reed1979} for quantum waves, of \cite{ramm1986} for classical waves. A generalized eigenfunction expansion
in terms of Lippman-Schwinger eigenvectors of the elastodynamic equation may also be found in the work of \cite{budreck1991}. Making
use of the  eigenvectors of the operator $L^0 + V(\omega_0^2)$ introduced in equation  (\ref{lippman}), we 
expand  Green's function $G$ as follows:
\begin{equation}
 G(\omega_0) = \sum_{\gamma} \iint_{4 \pi} d^2\hat{k} \int\limits^{+\infty}_0 
 \dfrac{ | \psi^{\gamma}(k\boldsymbol{\hat{k}})   \rangle \langle \psi^{\gamma}(k \boldsymbol{\hat{k}}) | }
{\rho_0 (\omega_0^2 -c^2_{\gamma}k^2 + i \omega_0/\tau)} k^2 dk  .
\label{G+}
\end{equation}
In the  case of quantum waves, this decomposition is correct if the scattering potential does not create bound states.
Using the formal definition of the $T^{(\omega_0^2)}$-operator (\ref{tmatop}) and the spectral representation (\ref{G+}), we find in the limit
($\tau \to \infty$):
\begin{multline}
\dfrac{1}{2i} \left( T^0_{\alpha \beta} (\boldsymbol{k}_a, \boldsymbol{k}_b)  - T^0_{\beta \alpha} (\boldsymbol{k}_b, \boldsymbol{k}_a)^* \right)= 
  \dfrac{1}{2i}\left(  \langle \alpha, \boldsymbol{k}_a | V(\omega_0^2)  | \beta, \boldsymbol{k}_b   \rangle   -  \langle \alpha ,\boldsymbol{k}_a | V(\omega_0^2)  | \beta ,\boldsymbol{k}_b   \rangle   \right)  +     
\\  \dfrac{1}{2i}\left( \langle \alpha, \boldsymbol{k}_a | V(\omega_0^2)  G(\omega_0)    V(\omega_0^2)  | \beta, \boldsymbol{k}_b  \rangle - 
       \langle \alpha, \boldsymbol{k}_a | V(\omega_0^2)  G(\omega_0)^{\dagger}    V(\omega_0^2)  | \beta, \boldsymbol{k}_b  \rangle   \right)  \\ 
 = -\dfrac{\omega_0}{\rho_0 \tau} \sum_{\gamma}   \iint_{4 \pi} d^2\hat{k} \int\limits^{+\infty}_0 k^2 \langle \alpha, \boldsymbol{k}_a | V(\omega_0^2)  
\dfrac{| \psi^{\gamma}(k\boldsymbol{\hat{k}})   \rangle \langle \psi^{\gamma}(k \boldsymbol{\hat{k}}) | }{\left[(\omega^2_0 - c^2_{\gamma}k^2)^2 +
 \omega_0^2/\tau^2) \right]} V(\omega_0^2)  | \beta, \boldsymbol{k}_b  \rangle dk \\
    =  -\dfrac{\pi}{2 \rho_0} \sum_{\gamma}   \iint_{4 \pi} d^2\hat{k} \int\limits^{+\infty}_0   \dfrac{k^2}{c_{\gamma} \omega_0}  
\left(\delta(k -\omega_0/c_{\gamma}) +
 \delta( k + \omega_0/c_{\gamma})  \right)  \\   \times \langle \alpha, \boldsymbol{k}_a | T^{\omega_0^2}(c_{\gamma} k) |\gamma, k\boldsymbol{\hat{k}} \rangle
                \langle \gamma, k\boldsymbol{\hat{k}} | T^{\omega_0^2}(c_{\gamma} k) | \beta, \boldsymbol{k}_b  \rangle    dk  \\
=   -\dfrac{\pi \omega_0}{2 \rho_0}  \sum_{\gamma} \dfrac{1}{c_{\gamma}^3}   \iint_{4 \pi} d^2\hat{k} 
T^{\omega_0^2}_{\alpha \gamma} \left(\boldsymbol{k}_a, \dfrac{\omega_0}{c_{\gamma}} \mathbf{\hat{k}} ;c^{\gamma}k =\omega_0 \right)  
T^{\omega_0^2}_{\beta \gamma} \left(\boldsymbol{k}_b, \dfrac{\omega_0}{c_{\gamma}} \mathbf{\hat{k}} ;c^{\gamma}k = \omega_0 \right)^*   \\
 =     -\dfrac{\pi \omega_0}{2 \rho_0}  \sum_{\gamma} \dfrac{1}{c_{\gamma}^3}   \iint_{4 \pi} d^2\hat{k} 
T^0_{ \gamma \alpha} \left( \dfrac{\omega_0}{c_{\gamma}} \mathbf{\hat{k}}, \boldsymbol{k}_a \right)^*  
T^0_{ \gamma \beta} \left( \dfrac{\omega_0}{c_{\gamma}} \mathbf{\hat{k}}, \boldsymbol{k}_b \right),
\label{proof}
\end{multline}
which is identical to equation (\ref{genoptthnoabs}). In the derivation of equation (\ref{proof}), we made use of the limit (\ref{funclim}) and of the reciprocity relations for the $T^{(\omega_0^2)}$-operator. In the third equality, we have made use of the
Lippman-Schwinger eigenvectors introduced in equation (\ref{lippman2}).
 It is to be noted that the previous results
are correct irrespective of the sign of $\omega_0$, which is important when dealing with classical waves.
\section{Application to a   point-scatterer model }
\label{point}
We now illustrate the process of Green's function reconstruction with a simple example of a 3-D heterogeneous medium
composed of a single delta-like scattering potential. 
To render multiple-scattering calculations easier, it is often convenient to introduce a point-scatterer model
in the same spirit as in the scalar case \citep{margerin2010}.   Mathematically, point-scattering poses some serious
difficulties \citep{devries1998}.
Physically, the point scatterer model is obtained by regularizing the
Born series for a scattering potential of the form:
\begin{equation}
 V(\omega_0^2) = \gamma(\omega_0^2) \mathcal{I} | \mathbf{x}_0 \rangle \langle \mathbf{x}_0 | , 
\label{scatpot1}
\end{equation}
where   $\mathbf{x}_0$ is the location of the scattering center, 
and the physical significance of the parameter $\gamma(\omega_0^2)$ will be further discussed below. 
 The scattering potential $V(\omega_0^2)$ has to be understood as a tensor product between two operators. The symbol  $\mathcal{I}$ denotes the identity operator in polarization space. It simply leaves unchanged the vectorial part  of the wavefunction.  The symbol $| \mathbf{x}_0 \rangle \langle \mathbf{x}_0 |$ is a projection operator which picks out the value of the wavefunction exactly at position $\mathbf{x}_0$, hence the name ``point-scatterer''. 
We will find that the scattering potential
$V(\omega_0^2)$  represents adequately a  small scatterer with perfectly correlated perturbations of the density and $P$ and $S$ wave velocities according to:
 $ \delta c_p/c_p = \delta c_s/c_s =- \delta \rho/2\rho,  $ ($\delta \lambda/\lambda = \delta \mu/\mu=0$).  No assumption on the smallness
of the perturbation is made.
In the representation (2), the scattering potential writes
\begin{equation}
 V_{ij}(\mathbf{k},\mathbf{k}';\omega_0^2) = \frac{\gamma(\omega_0^2)}{(2 \pi)^3} e^{i\mathbf{x}_0\cdot(\mathbf{k}'-\mathbf{k})} \delta_{ij}.
\label{scatpot2}
\end{equation}
To evaluate the scattering properties of such a perturbation, we make use of the formal Born series for the $T$-operator of elastic waves: 
\begin{equation}
 T^0 =  V(\omega_0^2) + V(\omega_0^2)G_0(\omega_0)V(\omega_0^2) + V(\omega_0^2)G_0(\omega_0)V(\omega_0^2)G_0(\omega_0)V(\omega_0^2) + ... 
\label{bornseries}
\end{equation}
and evaluate the  matrix elements in the representation (2) 
\begin{multline}
\langle \mathbf{k},i |  T^0 |j, \mathbf{k}' \rangle =   V_{ij}(\mathbf{k},\mathbf{k}';\omega_0^2) +
  \sum_{i_1} \iiint_{\mathbb{R}^3}   \langle \mathbf{k},i |V(\omega_0^2) | i_1, \mathbf{k}_1 \rangle \langle i_1 ,\mathbf{k}_1|
    G^0(\omega_0)V(\omega_0^2)|j, \mathbf{k}' \rangle  d^3k_1  +  \\
     \sum_{i_1,i_2}   \iiint_{\mathbb{R}^6}  \langle \mathbf{k} ,i  |V(\omega_0^2)  | i_1, \mathbf{k}_1 \rangle \langle \mathbf{k}_1 ,i_1| G^0(\omega_0)V(\omega_0^2) | i_2, \mathbf{k}_2 \rangle \langle i_2 , \mathbf{k}_2| G^0(\omega_0)V(\omega_0^2) | j , \mathbf{k}' \rangle    d^3k_1 d^3k_2  \\
   + \cdots
\label{bornseries2}
\end{multline}
The need for regularization stems from the fact that repeated visits
of the same scatterer involve evaluations of the  free space Green's function for coincident source and receiver, which clearly diverges.
We now show that for a scattering potential of the form (\ref{scatpot1}), the Born series (\ref{bornseries}) can be summed analytically after
the return Green's function is made finite.
To make sense of expression (\ref{bornseries2}), we need to  evaluate the matrix elements of the operator  $G_0(\omega_0)V(\omega_0^2)$:
\begin{equation}
\begin{split}
\langle \mathbf{k} ,i | G^0(\omega_0) V(\omega_0) | j, \mathbf{k}' \rangle = & 
 \sum_{i_1} \iiint_{\mathbb{R}^3}  \langle \mathbf{k} ,i | G^0(\omega_0)  | i_1, \mathbf{k}_1 \rangle 
     \langle i_1 ,\mathbf{k}_1| V(\omega_0^2) | j, \mathbf{k}' \rangle  d^3k_1 \\  
       = & \frac{\gamma(\omega_0^2) e^{i\mathbf{x}_0 \cdot(\mathbf{k}'-\mathbf{k})}}{(2 \pi)^3} \\
      & \times  \left( 
\dfrac{  \hat{k}_i \hat{k}_j }{\rho_0(\omega_0^2 - (c_{p} k)^2 +i \omega_0/\tau)} 
+ \frac{  \delta_{ij} -  \hat{k}_i \hat{k}_j }{\rho_0(\omega_0^2 - (c_{s} k)^2 +i \omega_0/\tau)} \right),
\label{g0v}
\end{split}
\end{equation}
where we have made use of equations (\ref{g0k}) and (\ref{scatpot2}).
Reporting expression  (\ref{g0v}) in the first integral of Equation (\ref{bornseries2}) we obtain:
\begin{equation}
\begin{split}
I_1 = & \frac{\gamma(\omega_0^2) e^{i\mathbf{x}_0 \cdot(\mathbf{k}'-\mathbf{k})}}{(2 \pi)^3} 
  \iiint_{\mathbb{R}^3} \left( \frac{   \mathbf{\hat{k}}_1 \mathbf{\hat{k}}_1}{\rho_0(\omega_0^2 - (c_{p} k_1)^2 +i \omega_0/\tau)}  
+  \frac{\mathcal{I} - \mathbf{\hat{k}}_1 \mathbf{\hat{k}}_1    }{\rho_0(\omega_0^2 - c_{s}^2 k_1^2 +i \omega_0/\tau)} \right) d^3 k_1  \\
       =&   \frac{\gamma(\omega_0^2)  e^{i \mathbf{x}_0(\mathbf{k}' -\mathbf{k} )} }{(2 \pi)^3}   \frac{ \mathcal{I} 4 \pi \gamma(\omega_0^2)}{3 (2 \pi)^3}  \left( 
 \int\limits_{0}^{\infty} \frac{ k_1^2 dk_1}{\rho_0(\omega_0^2 - c_{p}^2k_1^2 +i\omega_0/\tau)}  + 2 
 \int\limits_{0}^{\infty} \frac{ k_1^2 dk_1}{\rho_0(\omega_0^2 - c_{s}^2k_1^2 +i\omega_0/\tau)} \right)  ,
\label{I1}
\end{split}
\end{equation}
where we have used:
\begin{equation}
 \iint_{4 \pi} \mathbf{\hat{k}}_1 \mathbf{\hat{k}}_1   d^2\hat{k}_1 = \dfrac{4 \pi \mathcal{I}}{3} .
\end{equation}
Similarly, for the second integral of equation (\ref{bornseries}), one obtains:
\begin{equation}
\begin{split}
 I_2 = 
\frac{\gamma(\omega_0^2)}{(2 \pi)^3}  e^{i \mathbf{x}_0(\mathbf{k'} -\mathbf{k} )}  
  \left[ \frac{ \mathcal{I} 4 \pi \gamma(\omega_0^2)}{3 (2 \pi)^3}   \left( 
 \int\limits_{0}^{\infty} \frac{ k_1^2 dk_1}{\rho_0(\omega_0^2 - c_{p}^2k_1^2 +i\omega_0/\tau)}  + 2 
 \int\limits_{0}^{\infty} \frac{ k_1^2 dk_1}{\rho_0(\omega_0^2 - c_{s}^2k_1^2 +i\omega_0/\tau)} \right)\right]^2  .
\end{split}
\label{I2}
\end{equation}
From equations (\ref{I1})-(\ref{I2}), we infer that the Born series is in fact a simple geometric series.
The last integrals in (\ref{I1})-(\ref{I2}) diverge and need to be regularized. 
We proceed as in \cite{vanrossum1999}, by introducing a large momentum cut-off 
to avoid divergence of the real part of  Green's function. Let us illustrate this procedure
for the $P$ modes. We rewrite the first integral on the right-hand side of equation (\ref{I2}) as follows:
\begin{equation}
\begin{split}
J = & \frac{1}{\rho_0 c_{p}^2} \int\limits_0^{\infty} \frac{k_1^2 dk_1}{k^2_{p} - k_1^2 +i k_p/l_p } \\
  = & \frac{1}{\rho_0 c_{p}^2} \int\limits_0^{\infty} k_1^2 dk_1
 \left[ \frac{1}{(k_{p}^2 -  k_1^2 +i k_p/l_p)}  + \frac{1}{k_1^2} \right]
    - \frac{1}{\rho_0 c_{p}^2} \int\limits_0^{\infty} dk_1 ,
\end{split}
\end{equation}
where $k_p = \omega_0/c_p$ and $l_p = c_p \tau $. Note that  terms of order $1/l_p^2$ have been neglected in the numerator, which
implies that absorption is sufficiently small.
In the first integral, we have subtracted out the divergent part. In the second integral, we introduce
a large momentum cut-off with a scale length $a$ to obtain:
\begin{equation}
\begin{split}
J \approx & \frac{1}{\rho_0 c_{p}^2} \int\limits_0^{\infty} \frac{(k_p + i/2l_p)^2 dk_1}{k^2_{p} - k_1^2 +i k_p/l_p }
          -  \frac{1}{\rho_0 c_{p}^2} \int\limits_0^{1/a} dk_1  \\
  \approx &  -\frac{1}{\rho_0 c_{p}^2} \left( \frac1a  -\dfrac{\pi}{4 l_p} + i \frac{\pi k_{p}}{2} \right),
\label{regular}
\end{split}
\end{equation}
where $k_p = \omega_0/c_p$.
Combining equations  (\ref{I1})-(\ref{I2}) and (\ref{regular}), one may rewrite the formal series expression of  the $T^0$-matrix  
given by (\ref{bornseries}) in the form: 
\begin{equation}
T^0_{ij}(\mathbf{k},\mathbf{k}') = \dfrac{e^{i\mathbf{x}_0(\mathbf{k}'-\mathbf{k})}t(\omega_0) \delta_{ij}}{(2\pi)^3} ,
\label{top}
\end{equation}
where the scalar $t(\omega_0)$-matrix  can be expressed as:
\begin{equation}
t(\omega_0)=  \dfrac{\gamma(\omega_0^2)}{1- \gamma(\omega_0^2) U(\omega_0)},
\label{tmat}
\end{equation}
and  we have introduced an effective potential:
\begin{align}
U(\omega_0) = & -\frac{1}{6 \rho_0 \pi^2}\left[ \kappa   + i\pi\left(\frac{\omega_0}{2c_p^3} + \frac{\omega_0}{c_s^3} \right)  \right]  \\    
\kappa = & \frac{1}{a}\left(\frac{2}{c_s^2} + \frac{1}{c_p^2} \right)   - \dfrac{\pi}{2 \tau} \left( \frac{1}{2c_p^3} + \frac{1}{c_s^3}   \right) 
\end{align}
where the length $a$ stems from the large momentum cut-off. The result (\ref{tmat}) is valid irrespective of the sign of $\omega_0$.
Equations (\ref{top})-(\ref{tmat}) offer a generalization of the point-scattering model to vector elastic waves.
The regularization procedure is not unique as illustrated by \cite{devries1998} for scalar and electromagnetic waves.
 The most important point  is the preservation of the local imaginary part of the free space Green's function:
 \begin{equation}
 \operatorname{Im} G^0_{ij}(\mathbf{x},\mathbf{x}) =    -\dfrac{\delta_{ij}}{4 \pi \rho_0}\left(\frac{\omega}{3 c_p^3} + \frac{2\omega}{3c_s^3 } \right), 
\label{imgxx}
 \end{equation}
 which is proportional to  the density of states, i.e. the number of modes per unit volume and frequency.
We note  that the imaginary part of the free space Green's function at $(\mathbf{x},\mathbf{x})$ is an equipartition mixture of $P$ and $S$ waves 
and does not depend on the absorption time $\tau$.
The result (\ref{tmat}) for the $t$-matrix is valid for a small, finite amount of uniform absorption inside the medium. Note that
because the absorbing properties of the scatterer match those of the embedding matrix the scattering potential is real.

To analyze the properties of our point scatterer, we make use of the relation (\ref{scatamp}) providing the relation
between the scattering amplitude and the $T^0$-operator. The scattering pattern is obtained by 
sandwiching the $T^0$-operator  between two plane waves, i.e. we take the matrix elements of $T^0$
in representation (3).
 \begin{equation}
   f^{\alpha \leftarrow \beta}(\mathbf{\hat{k}}_a;\mathbf{\hat{k}}_b) \propto   \langle \mathbf{k}_a, \alpha | T^0 | \mathbf{k}_{b},\beta \rangle \\
                                  \propto   \langle \mathbf{p}^{\alpha} |\mathcal{I} | \mathbf{\hat{q}}^{\beta}  \rangle
                            =  \mathbf{\hat{p}}^{\alpha}\cdot  \mathbf{\hat{q}}^{\beta} ,     
 \end{equation} 
where $ \mathbf{\hat{p}}^{\alpha}, \mathbf{\hat{q}}^{\beta} $ denote the polarization vector of the scattered and incident wave, respectively. Let us specialize the last result by introducing a spherical coordinate system where the polar angle $\theta$ is measured from the axis pointing in the direction of the incident wave with wavevector $\mathbf{k}_b$. For the different scattering type, we obtain the following angular dependence:
(1) $P$ to $P$: $\cos \theta$; (2) $P$ to $SV$ and $SV$ to $P$: $\sin \theta$; 
(3) $SV$ to $SV$: $\cos \theta$ , (4) $SH$ to $SH$: no angular dependence (in and out polarization vectors are parallel),
where $SV$ and $SH$ denote transverse polarizations in and perpendicular to the scattering plane, respectively. All other coupling terms are zero.
 These results agree with the scattering pattern for a pure density perturbation
  (constant Lam\'e parameters) in the low-frequency or Rayleigh regime. 
 The scattering pattern is anisotropic  but non-preferential, i.e., there is equal amount of forward and backward scattering. 
This property is typical of a scatterer whose size is much smaller than the wavelength.
There is no mode conversion in the forward direction as required by symmetry consideration. 
Reporting expression (\ref{top}) in the  usual form of the optical theorem (\ref{clasoptth2}), we find the following  relations for
the scattering cross-sections in terms of the $t$-matrix:
\begin{subequations}
\begin{align}
\sigma_p = & -\dfrac{1}{\rho_0 c_p} \text{Im} \dfrac{t}{\omega_0}  \label{sigmap}  \\
\sigma_s = & -\dfrac{1}{\rho_0 c_s} \text{Im} \dfrac{t}{\omega_0} ,
\end{align}
\end{subequations}
which implies:
\begin{equation}
 c_p \sigma_p  = c_s \sigma_s.
\label{cpsigmap}
\end{equation}
The relation (\ref{cpsigmap}) is valid for a point scatterer but not generally true.
Finally, let us identify the parameter $\gamma(\omega_0^2)$ with $-\delta \rho \omega_0^2 \mathcal{V}$, with $\mathcal{V}$ the volume of the scatterer and
$\delta \rho$ the density contrast between the scatterer and the matrix. By substituting expression
(\ref{tmat}) into expression (\ref{sigmap}), and taking  the limit $\omega_0 \to 0$, we find:
\begin{equation}
 \sigma_p = \dfrac{2\mathcal{V}}{9}\left(\dfrac{1}{2 c_p^3} + \dfrac{1}{c_s^3}   \right)  \left( \dfrac{\delta \rho}{\rho_0} \right)^2 \dfrac{\omega_0^4 r^3}{c_p},
\end{equation}
with $r$ the scatterer radius. These results concur with previous analytical solutions of the scattering problem for a spherical
inclusion in the Rayleigh regime  \citep{korneev1993}.
We have therefore defined a physically meaningful point-scatterer for elastic waves.
It is important to note that relation ($\ref{cpsigmap}$) pertains to a localized perturbation of the density at \emph{fixed} Lam\'e parameters
$\lambda$ and $\mu$.

The final step of our investigation of the point scatterer concerns the verification of the complete identity
(\ref{genoptthabs}). On the right-hand side, we find:
\begin{equation}
   \dfrac{1}{2i} \left(  T^0_{\alpha \beta}(\mathbf{k}_a,\mathbf{k}_b) - T^0_{\beta \alpha}(\mathbf{k}_b,\mathbf{k}_a)^*   \right)   = 
  \dfrac{ \mathbf{\hat{p}}^{\alpha} \cdot   \mathbf{\hat{q}}^{\beta} e^{i\mathbf{x}_0 \cdot(\mathbf{k}_b -\mathbf{k}_a)} \text{Im}t(\omega_0)}{(2 \pi)^3 }    
\end{equation}
The left-hand side can be calculated as follows:
\begin{multline}
    -\dfrac{\omega_0}{\rho_0 \tau} \sum_{\gamma} \iiint  \dfrac{ T^0_{\gamma \beta}(\mathbf{k}, \mathbf{k}_b  ) 
       T^0_{\gamma \alpha}( \mathbf{k},  \mathbf{k}_a  )^* }
       {  \left( \omega_0^2 - (c_{\gamma}k)^2\right)^2 + \omega_0^2/\tau^2   }    d^3k =  \\   -\dfrac{\omega_0}{\rho_0 \tau (2 \pi)^6} \sum_{\gamma} \iiint 
 \dfrac{ |t(\omega_0)|^2  e^{i \mathbf{x}_0 \cdot (\mathbf{k}_b - \mathbf{k}_a)}  
    \langle \mathbf{\hat{p}}^{\alpha} | \gamma, \mathbf{k} \rangle \langle  \gamma, \mathbf{k} |  \mathbf{\hat{q}}^{\beta} \rangle       } 
       {  \left( \omega_0^2 - (c_{\gamma}k)^2\right)^2 + \omega_0^2/\tau^2   }    d^3k   \\
 =    -\dfrac{4 \pi \omega_0  |t(\omega_0)|^2 \mathbf{\hat{p}}^{\alpha} \cdot   \mathbf{\hat{q}}^{\beta}  }{3  \rho_0 \tau (2 \pi)^6}  
  \sum_{\gamma} \int_0^{+\infty} \dfrac{k^2 dk}{ \left( \omega_0^2 - (c_{\gamma}k)^2\right)^2 + \omega_0^2/\tau^2 }             \\
      =  \dfrac{-\omega_0 |t(\omega_0)|^2  \mathbf{\hat{p}}^{\alpha} \cdot   \mathbf{\hat{q}}^{\beta}}{(2 \pi)^3 6 \pi \rho_0} \left(\dfrac{1}{c_s^3} + \dfrac{1}{2c_p^3}  \right),
\end{multline}
a result valid in both absorbing (finite $\tau$) and non-absorbing media ($\tau \to \infty$).
From equation (\ref{imgxx}), we deduce a relation analog to the scalar case \citep{vanrossum1999}:
\begin{equation}
\text{Im} t(\omega_0) \mathcal{I} = |t(\omega_0)|^2 \text{Im} G^0(\mathbf{x},\mathbf{x};\omega_0).
\label{imt}
\end{equation}
A straightforward calculation shows that   relation (\ref{imt}) is indeed verified by the $t$-matrix obtained from the  regularization procedure.
With this simple example we have therefore illustrated the relation between the reconstruction of the elastic Green's function and the conservation
of energy in scattering. These results can be generalized to more complex multiple-scattering media. Such developments will be presented elsewhere.
\section{Discussion and Conclusion}
\label{conclusion}
\begin{figure}
\centering
\includegraphics[width=0.7\linewidth]{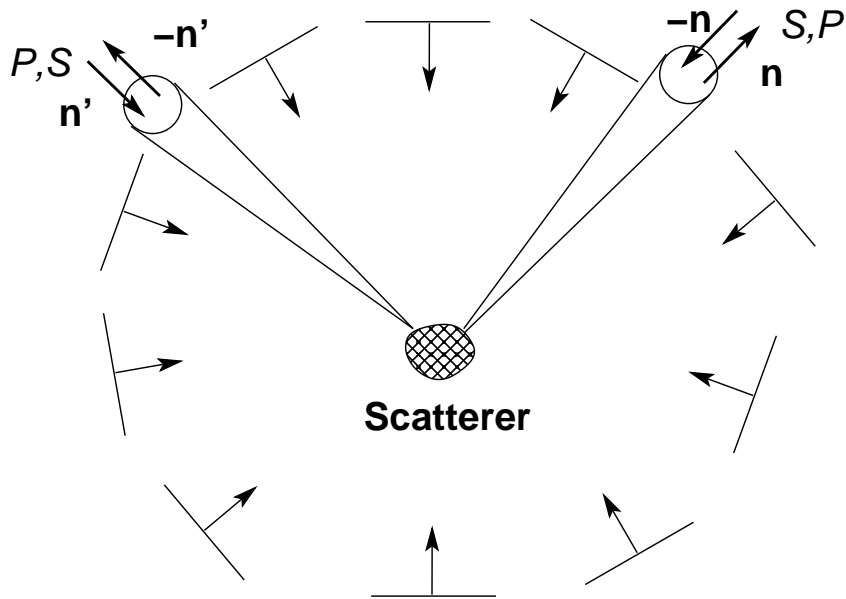}
\caption{Energy balance of a random wavefield in a homogeneous isotropic elastic medium containing a scatterer.
 The scatterer is illuminated by uncorrelated plane $P$ and $S$ waves
coming from all possible space directions. The total energy carried by $P$ waves increases because an $S$ wave propagating in direction
$\mathbf{n'} $ can be converted into a  $P$  wave propagating in direction $\mathbf{n}$. The reciprocal situation shows a $P$ wave propagating
in direction $-\mathbf{n}$ and converted into an $S$ wave propagating in direction $-\mathbf{n'}$. This scattering process
  entails a decrease of the total $P$ wave  energy. For an incident wavefield at equipartition the two contributions cancel exactly.
The same argument applies to the total $S$ energy.}
\label{balance}
\end{figure}
The reconstruction of   Green's function of a homogeneous elastic medium from random noise sources  was previously studied by
 \cite{SanchezSesma.Campillo.2006}. These authors used  a decomposition of the random wavefield
into a sum of uncorrelated  plane $P$ and $S$ waves at equipartition. In this paper, we gave a precise mathematical  formulation of the equipartition 
condition  in equation (\ref{equipartition}). Using the Dirac calculus, we can show the complete equivalence between the equipartition state and an excitation
of the system by randomly oriented and homogeneously distributed uncorrelated forces. 
 Noting the relations:
\begin{subequations}
\begin{align}
   f_l(\mathbf{x}_1;\omega_1) f_k(\mathbf{x}_0;\omega_0)^*  =  &   \langle l, \mathbf{x}_1  |f(\omega_1) \rangle \langle f(\omega_0) | k, \mathbf{x}_0 \rangle \\
    \tilde{f}_{\beta}(\mathbf{k}_1;\omega_1) \tilde{f}_{\alpha}(\mathbf{k}_0;\omega_0)^*  =   &  \langle \beta, \mathbf{k}_1  |f(\omega_1) \rangle \langle f(\omega_0) | \alpha, \mathbf{k}_0 \rangle,
 \end{align}
\end{subequations}
we obtain:
\begin{equation}
\begin{split}
   \llangle \tilde{f}_{\beta}(\mathbf{k}_1;\omega_1)  \tilde{f}_{\alpha}(\mathbf{k}_0;\omega_0)^* \rrangle = &     \\
& \text{\hspace*{-2cm} } \sum_{k,l} \iiint_{\mathbb{R}^6}  \llangle ( \langle \beta, \mathbf{k}_1  |l, \mathbf{x}_1 \rangle  \langle l, \mathbf{x}_1  |f(\omega_1) \rangle
       \langle f(\omega_0) | k, \mathbf{x}_0 \rangle \langle  k, \mathbf{x}_0  | \alpha, \mathbf{k}_0 \rangle) \rrangle  d^3x_0 d^3x_1 
  \\
  = & 
  2 \pi S(\omega_0)\delta(\omega_0-\omega_1) \\
    & \times\sum_{k,l} \iiint_{\mathbb{R}^6} \delta(\mathbf{x}_0 -\mathbf{x}_1)  \delta_{kl} ( \langle \beta, \mathbf{k}_1  |k, \mathbf{x}_0 \rangle 
     \langle  l, \mathbf{x}_1  | \alpha, \mathbf{k}_0 \rangle)  d^3x_0 d^3x_1 \\
 = & 2\pi S(\omega_0) \delta(\omega_0 - \omega_1) \sum_{k} \iiint_{\mathbb{R}^3}  \langle \beta, \mathbf{k}_1  |k, \mathbf{x}_0 \rangle \langle k, \mathbf{x}_0| \alpha, \mathbf{k}_0\rangle d^3x_0 \\
 = & 2 \pi S(\omega_0) \delta(\omega_0 -\omega_1)\delta_{\alpha \beta} \delta(\mathbf{k}_0 - \mathbf{k}_1) 
\label{equivalence}
 \end{split}
\end{equation}
Equation (\ref{equivalence}), demonstrates that from the point of view
of  correlations, a white noise distribution of random forces and a wavefield at equipartition are  equivalent. 
We may also illustrate this point with physical arguments. 
First we note that the  radiation of energy by a single force can be evaluated from the far-field solutions. For the $P$-wave radiation, we multiply $r^2 \rho_0 \omega_0^2 c_p$ by the squared modulus  of  Green's function $g^p(\mathbf{r}; \omega_0)$  as given by Equation (\ref{gprom}). The procedure is the same for the $S$ wave radiation, except for the substitution $c_p \rightarrow c_s$. It follows that the ratio between the total $S$- and $P$- energy radiated by a force
 is $2(c_p/c_s)^3$, which  rigorously coincides with the ratio of the equipartition state. A homogeneous distribution of randomly
oriented forces will therefore generate $P$ and $S$ waves propagating in all possible directions with a  local energy density of $P$ and $S$
waves satisfying the equipartition relation $E^S/E^P=2c_p^3/c_s^3$. Thus we find that the force distribution (\ref{randomforce1}) guarantees the existence
of the equipartition state and the reconstruction of  Green's function.
 It is important to note that the model of absorption that we have developed implies equality of the
intrinsic quality factor between $P$ and $S$ waves. This assumption may not  always be fulfilled in practice.
If the absorption times of  $P$ and $S$ waves are different,  the $P$-to-$S$ energy ratio differs from that of the equipartition state and Green's function 
reconstruction will be imperfect.

The requirement of equipartition for the successful reconstruction of  Green's function of a heterogeneous medium has been widely noted.
\cite{derode2003} have studied numerically and theoretically the reconstruction of Green's function of a heterogeneous, open medium  illuminated
by either random or deterministic sources. They gave a criterion based on time-reversal arguments to discuss the optimal arrangement of the sources. 
The reconstruction of  Green's function of a heterogeneous medium illuminated by an incident random wavefield at equipartition
has been considered theoretically by \cite{weaver2004} for scalar waves.  
\cite{sanchezsesma2006b}  have demonstrated  that 
when a cylindrical scatterer is illuminated by random plane $P$ and $S$ waves at equipartition in 2-D, the cross-correlation of the wavefield
allows the reconstruction of the complete Green's function, including the waves scattered by the cylinder.
  In the case of a distribution of sources on a remote surface enclosing the receivers and a  3-D isotropic and homogeneous  heterogeneity, 
 \cite{lu2011}  showed that the wavefield incident on the scatterer must obey an equipartition relation. Their derivation supposes that the medium is free from absorption.
 In this work, we have generalized
these results to an arbitrary, localized heterogeneity and have given an application to a point-scatterer. 
To make the link with previous investigations,  we now consider the impact of a localized scatterer on the equipartition state. 
In Figure \ref{balance}, we present a detailed balance of the energy carried by each wave mode to show that equipartition is globally
preserved in a scattering system. Let us consider a bundle of plane $P$ waves coming from space direction $\mathbf{n'}$ with
mean squared amplitude $\llangle |A^P(\mathbf{n'}) |^2 \rrangle$. Upon scattering, the $P$ waves will be mode converted into $S$ waves
propagating  into direction $\mathbf{n}$. Reciprocally, we may consider incident $S$ waves propagating in direction $-\mathbf{n}$ with
mean-squared amplitude $\llangle |A^S(\mathbf{-n}) |^2 \rrangle$ and mode converted
 into $P$ waves propagating in direction $-\mathbf{n'}$ . Let us now consider the balance of  energy
carried by $P$ waves in the scattering process:
\begin{equation}
\begin{split}
 \dfrac{d E^P}{dt} = & - \rho_0 \omega_0^2 c_s \iint_{4\pi \times 4\pi} d^2n  \llangle |A^P(\mathbf{n'}) |^2 \rrangle (|f^{S_1 \leftarrow P }(\mathbf{n},\mathbf{n}')|^2 + 
|f^{S_2 \leftarrow P }(\mathbf{n},\mathbf{n}')|^2   )d^2n' \\
     & + \rho_0 \omega_0^2 c_p \iint_{4\pi \times 4\pi} d^2n   (  \llangle |A^{S_1}(-\mathbf{n}) |^2 \rrangle |f^{P \leftarrow S_1 }(-\mathbf{n'},-\mathbf{n})|^2   \\
   &   \text{\hspace*{2.2cm}} \llangle |A^{S_2}(-\mathbf{n}) |^2 \rrangle  |f^{P \leftarrow S_2 }(-\mathbf{n'},-\mathbf{n})|^2   )d^2n',
\label{eqbalance}
\end{split}
\end{equation}
where $E^P$ denotes the total energy carried by $P$ waves.
In equation (\ref{eqbalance}), the integrals are performed over all possible space directions $\mathbf{n}$ and $\mathbf{n}'$. 
The first term represents the rate  of energy conversion from $P$ waves to $S$ waves in the scattering process.  This contributes
to a decrease to the total $P$ energy, hence the minus sign. The second term corresponds to the reciprocal process where $P$ energy increases because of the $S$ to $P$ mode conversions. For the description of the shear wave motion, we have chosen a basis of linear polarizations $\left\{S_1,S_2\right\}$ as illustrated
in Figure \ref{notation}. Other  bases such as left-handed and right-handed circular polarizations are popular in optics.  
  For a wavefield at equipartition the mean squared amplitudes are independent of the vectors $\mathbf{n}$
 and $\mathbf{n'}$ and satisfy the relations:
$ c_s^3 \llangle |A^{S_1}|^2 \rrangle = c_p^3 \llangle |A^P|^2 \rrangle $,  $\llangle |A^{S_1}|^2 \rrangle = \llangle |A^{S_2}|^2 \rrangle $.
Because of the fundamental reciprocity relations of conversion scattering derived
in equation (\ref{reciprocityf}), we find that 
the total energy carried by $P$ waves is constant, i.e. $dE^P/dt=0$. This implies that equipartition is globally preserved by the introduction
of a scatterer in the medium. This simple argument highlights again the link between Green's function reconstruction and equipartition.

In this paper, the link between symmetry relations in scattering and the reconstruction of Green's function from
random noise sources has been generalized to the case of elastic waves 
(see \cite{Snieder.2008,Snieder.Fleury.2010,margerin2010} for the acoustic case). Using the Dirac calculus, a number
of previously known symmetry and reciprocity relations for the   scattering amplitude  have been derived in a systematic way
 \citep{varatharajulu1977,dassios1987,aki1992}.
Based on the general Green's function reconstruction formula for stationary random noise sources, we have obtained a general form of optical 
theorem valid in both near and far-field, which incorporates  the effect of a finite amount of absorption. We also show that the generalized
optical  theorem can  be established from the scattering formalism in the limit of infinitesimal absorption. 
Thus, the correlation properties of random wavefields  offer an alternative
approach to study scattering phenomena in heterogeneous media.
The reconstruction of Green's function  in 3-D is illustrated with the simple case of a point-like heterogeneity embedded in an isotropic
matrix. Based on a regularization of the return free space Green's function, we are able to sum
the Born series for such a  defect. Usual results of the formal boundary value problem for a small scatterer with an arbitrary 
density contrast are recovered in the low frequency limit. The point scatterer satisfies the generalized optical theorem, which in turn implies the 
reconstruction of Green's function in the presence of random distribution of noise sources.  The point scattering model offers 
an attractive approach to tackle more complex multiple-scattering media. Such developments will be the topic of a separate publication.

\section*{Acknowledgments}
We would like to thank K. Wapenaar for his detailed review, and for his suggestions to improve the clarity of the manuscript. We thank an anonymous referee for drawing our attention
to recent references on Green's function retrieval. This work was supported by a joint program of the Japanese society for the promotion of science (Japan) and the Centre national de la recherche scientifique (France).


\end{document}